\documentclass[fleqn,aps,prb,twocolumn,showpacs,floatfix,longbibliography]{revtex4-2}
\usepackage{amsmath,amssymb,amsfonts,float,graphics,epsfig,epstopdf,color,verbatim,tabularx,bm,multirow,appendix,hyperref}

\usepackage{amsthm}
\theoremstyle{definition}

\usepackage{amsmath}
\usepackage{amssymb}
\usepackage{mathtools}
\usepackage{graphicx}
\usepackage{lipsum}
\usepackage{bm}
\usepackage{hyperref}
\usepackage{url}
\usepackage[utf8]{inputenc}
\usepackage{subfigure}
\usepackage{slashed,bbm}
\usepackage{graphics,psfrag,epsfig}
\usepackage{dsfont}
\usepackage{setspace}
\usepackage{wasysym}
\usepackage{slashed}
\usepackage{lipsum}
\usepackage{braket}
\usepackage{physics}
\usepackage{enumerate}%
\usepackage{makecell}
\usepackage{xcolor}

\definecolor{dark-red}{rgb}{0.4,0.15,0.15}
\definecolor{dark-blue}{rgb}{0.15,0.15,0.4}
\definecolor{medium-blue}{rgb}{0,0,0.5}
\hypersetup{
colorlinks, linkcolor={dark-blue},
citecolor={dark-blue}, urlcolor={medium-blue}
}
\usepackage{ulem}

\newcommand{\be}{\begin{equation}}
\newcommand{\ee}{\end{equation}}
\newcommand{\bea}{\begin{eqnarray}}
\newcommand{\eea}{\end{eqnarray}}

\renewcommand{\i}{\text{i}}

\begin{document}

\title{Local Reversibility and Divergent Markov Length in 1+1-D Directed Percolation}

 \author{Yu-Hsueh Chen}
 \affiliation{Department of Physics, University of California at San Diego, La Jolla, California 92093, USA}
 \author{Tarun Grover}
 \affiliation{Department of Physics, University of California at San Diego, La Jolla, California 92093, USA}

\begin{abstract}
Recent progress in open many-body quantum systems has highlighted the importance of the Markov length, the characteristic scale over which conditional correlations decay. It has been proposed that non-equilibrium phases of matter can be defined as equivalence classes of states connected by short-time evolution while maintaining a finite Markov length, a notion called local reversibility. A natural question is whether well-known classical models of non-equilibrium criticality fit within this framework. Here we investigate the Domany--Kinzel model --- which exhibits an active phase and an absorbing phase separated by a 1+1-D directed-percolation transition --- from this information-theoretic perspective. 
Using tensor network simulations, we provide  evidence for local reversibility within the active phase. Notably, the Markov length diverges upon approaching the critical point, unlike classical equilibrium transitions where Markov length is zero due to their Gibbs character. Correspondingly, the conditional mutual information exhibits scaling consistent with directed percolation universality. Further, we analytically study the case of 1+1-D compact directed percolation, where the Markov length diverges throughout the phase diagram due to spontaneous breaking of domain-wall parity symmetry from strong to weak. Nevertheless, the conditional mutual information continues to faithfully detect the corresponding phase transition.

\end{abstract}
\maketitle

\textbf{Introduction:}
Phases and phase transitions that cannot be described by a Gibbs state are ubiquitous, but our understanding of them is not as organized as that for Gibbs states. An important idea that has been fruitful in both equilibrium and out-of-equilibrium physics is to define a phase of matter as an equivalence relation between states under certain dynamical operations. For example, for pure gapped ground states, this equivalence relation corresponds to being connected to the ground state of a fixed-point Hamiltonian via a finite-depth local unitary circuit~\cite{chen2010local,hastings2005quasiadiabatic}. For mixed states, time evolution is governed by quantum channels, and therefore, it is natural to seek an equivalence relation defined via a finite-time evolution by local channels~\cite{hastings2011topological,konig2014generating,coser2019classification}. Importantly, since channels are not generically reversible, one requires that the states in the same `mixed-state phase of matter' are connected by a finite-time channel in either direction~\cite{coser2019classification,sang2024mixed}. To make the correspondence even closer with pure states, one may further require that the steady states in the same phase of matter are connected by a finite-depth channel in either direction that is locally reversible~\cite{sang2025stability, sang2025mixed}.  Despite the apparent generality of this definition, this conceptual framework has  been tested only in a few non-equilibrium settings, primarily those that appear in the context of quantum error correction~\cite{sang2024mixed, sang2025stability, sang2025mixed}. If these notions are indeed general, they should be widely useful, including in purely classical settings. Motivated by such considerations, in this work we will study a few well-known examples of \textit{classical} non-equilibrium phases and phase transitions from this modern perspective. In particular, we focus on non-equilibrium models with absorbing states, whose critical behavior falls within the directed percolation and closely related universality classes.

\begin{figure}
\centering
\includegraphics[width=\linewidth]{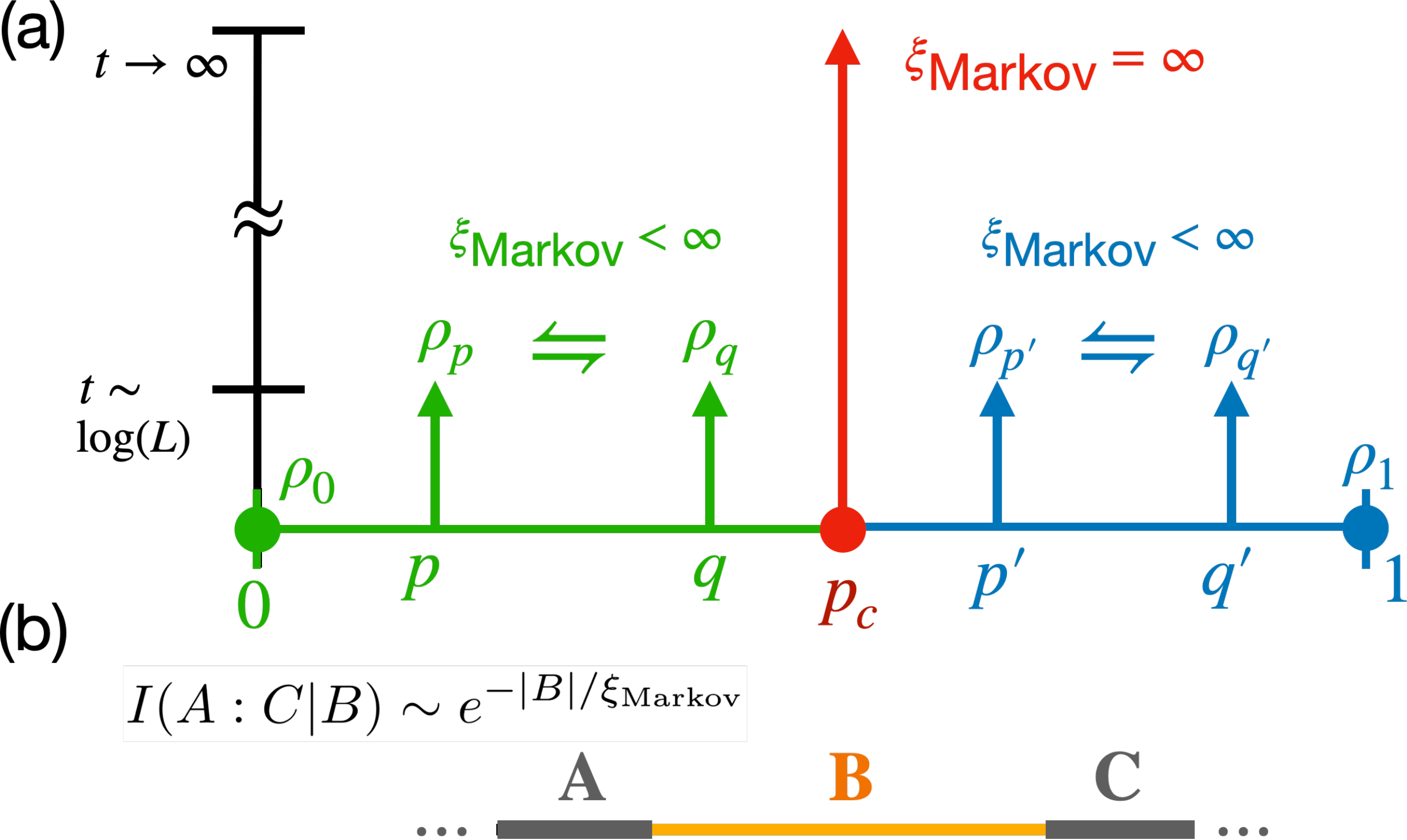}
\caption{
Summary of main results. (a) Starting from the non-absorbing steady state $\rho_0$ at $p=0$, the time-evolved state $\rho(t) = \mathcal{E}_p^t[\rho_0]$ in the active phase of the directed percolation ($p < p_c$) relaxes exponentially fast to the stationary state $\rho_p \equiv \mathcal{E}_p^\infty[\rho_0]$ ($L$ is the system size). Furthermore, the Markov length of $\rho(t)$ remains finite, implying local reversibility of the active phase.
In contrast, at the critical point, $p = p_c$, the state $\rho(t) = \mathcal{E}_{p_c}^t[\rho_0]$ converges only polynomially to $\rho_{p_c} \equiv \mathcal{E}_{p_c}^\infty[\rho_0]$, and the associated Markov length diverges.
 (b) The Markov length $\xi_{\text{Markov}}$ is defined via the decay length of the conditional mutual information, $I(A\!:\!C \mid B) \sim e^{-|B|/\xi_{\text{Markov}}}$ with the buffer size $|B|$. Similar situation holds (trivially) for the absorbing phase of the DP ($p > p_c$).
}
\label{fig:fig1}
\end{figure}

Directed percolation (DP) is a paradigmatic example of a non-equilibrium phase transition, relevant to a rather wide class of systems, ranging from reaction-diffusion processes to models of population spreading, and even to the onset of turbulence in certain fluid flows~\cite{schlogl1972chemical,cardy1980directed,grassberger1978reggeon,grassberger1979reggeon,janssen1981nonequilibrium,grassberger1981phase, kinzel1985phase,hinrichsen2000non,goldenfeld2017turbulence, tauber2014critical,cardy1996scaling}. In fact, it has been conjectured that under fairly general circumstances, all models with a unique absorbing state belong to the DP universality class~\cite{janssen1981nonequilibrium,grassberger1981phase}. A representative realization is provided by population dynamics, with a local density $\phi(x,t)$ governed by birth at rate $\tau$ and death at rate $g$ due to resource competition. A coarse-grained description of the time evolution of the population is given by $\partial_t{\phi} = \nabla^2 \phi + \tau \phi - g\phi^2/2 + \zeta$, where $\zeta$ is a noise term with zero mean, $\langle \zeta \rangle = 0$, and short-range spatio-temporal correlations $\langle \zeta(x,t) \zeta({x'},t') \rangle = g \phi(x,t) \delta({x} - {x'}) \delta(t - t')$. The fact that the noise correlations are proportional to the population density $\phi$ itself reflects that the fluctuations vanish when $\phi(x,t) = 0$ (death is irreversible). Therefore, a state with $\phi({x}) = 0 \,\, \forall \,{x}$ is an absorbing state, i.e., the system can never leave this state. For $\tau < \tau_c $ ($\tau_c$ is the critical birth rate), the absorbing state is the only steady state, while for  $\tau > \tau_c$, there exist both the absorbing steady state and a state with non-zero population/`activity'. These two regimes are separated by a continuous phase transition in the DP universality class~\cite{schlogl1972chemical,cardy1980directed,grassberger1978reggeon,grassberger1979reggeon,janssen1981nonequilibrium,kinzel1985phase,tauber2014critical,cardy1996scaling}.

Recent renewed interest in non-equilibrium phases and phase transitions, driven in part by efforts to develop and characterize non-equilibrium quantum memories~\cite{fan2023diagnostics,bao2023mixed,lee2023quantum, wang2025intrinsic, chen2023separability, li2024replica, wang2025analog, lessa2025higher,ellison2025toward,sohal2025noisy,balasubramanian2024local}, has sharpened the focus on defining precisely what constitutes a non-equilibrium phase of matter. In particular, it has been postulated that for a system size $L$, if a state $\rho_1$ can be transformed into a state $\rho_2$ via a locally reversible local quantum channel $E_{12}$ of depth poly(log($L$)) (``short-depth''), i.e., $E_{12}[\rho_1] = \rho_2$, then they belong to the same phase of matter~\cite{coser2019classification,sang2024mixed,sang2025stability,sang2025mixed}. Local reversibility is the condition that one can locally undo the effect of a  channel and therefore, the existence of the channel $E_{12}$ guarantees existence of a local, short-depth channel in the opposite direction, $E_{21}[\rho_2] = \rho_1$~\cite{sang2025stability,sang2025mixed}. The central quantity that ensures the existence of the channel $E_{12}$ (and hence $E_{21}$) is the existence of a finite Markov length $\xi_{\textrm{Markov}}$~\cite{sang2025stability}, which is defined via the conditional mutual information (CMI) $I(A:C|B) = S(AB) + S(BC) - S(B) - S(ABC) \sim e^{-|B|/\xi_{\textrm{Markov}}}$, see Fig.\ref{fig:fig1} for the geometry (here $S(X)$ is the von Neumann entropy for region $X$). The intuition is that a finite Markov length implies that all correlations between subregions $A$ and $C$ are mediated via the subregion $B$, and therefore the action of the channel $E_{12}$ in a small region can be undone via the Petz recovery map~\cite{ohya2004quantum,petz1986sufficient,junge2018universal,sang2025stability}, thereby allowing one to explicitly construct the reverse channel $E_{21}$. We note that a related but distinct notion of non-equilibrium stability requires that small perturbations of the dynamical rules yield steady states that can be connected to one another by a finite-time evolution~\cite{rakovszky2024defining, cubitt2015stability, liu2024dissipative}.

Notably, the Markov length $\xi_{\textrm{Markov}}$ is exactly zero for Gibbs states of local, commuting Hamiltonians~\cite{clifford1971markov, leifer2008quantum,brown2012quantum}, and it has also been recently shown to be finite for the Gibbs state of any local Hamiltonian~\cite{chen2025quantum}. Therefore, $\xi_{\textrm{Markov}}$ also characterizes how non-Gibbsian a non-equilibrium state is, a question that has attracted considerable attention~\cite{lebowitz1988pseudo,lebowitz1990statistical, van1993regularity,fernandez2001non}. Remarkably, it has been found that the phase transitions corresponding to error threshold in certain quantum error correcting codes have a \textit{divergent} Markov length~\cite{sang2025stability}.

\begin{figure}
\centering
\includegraphics[width=0.6\linewidth]{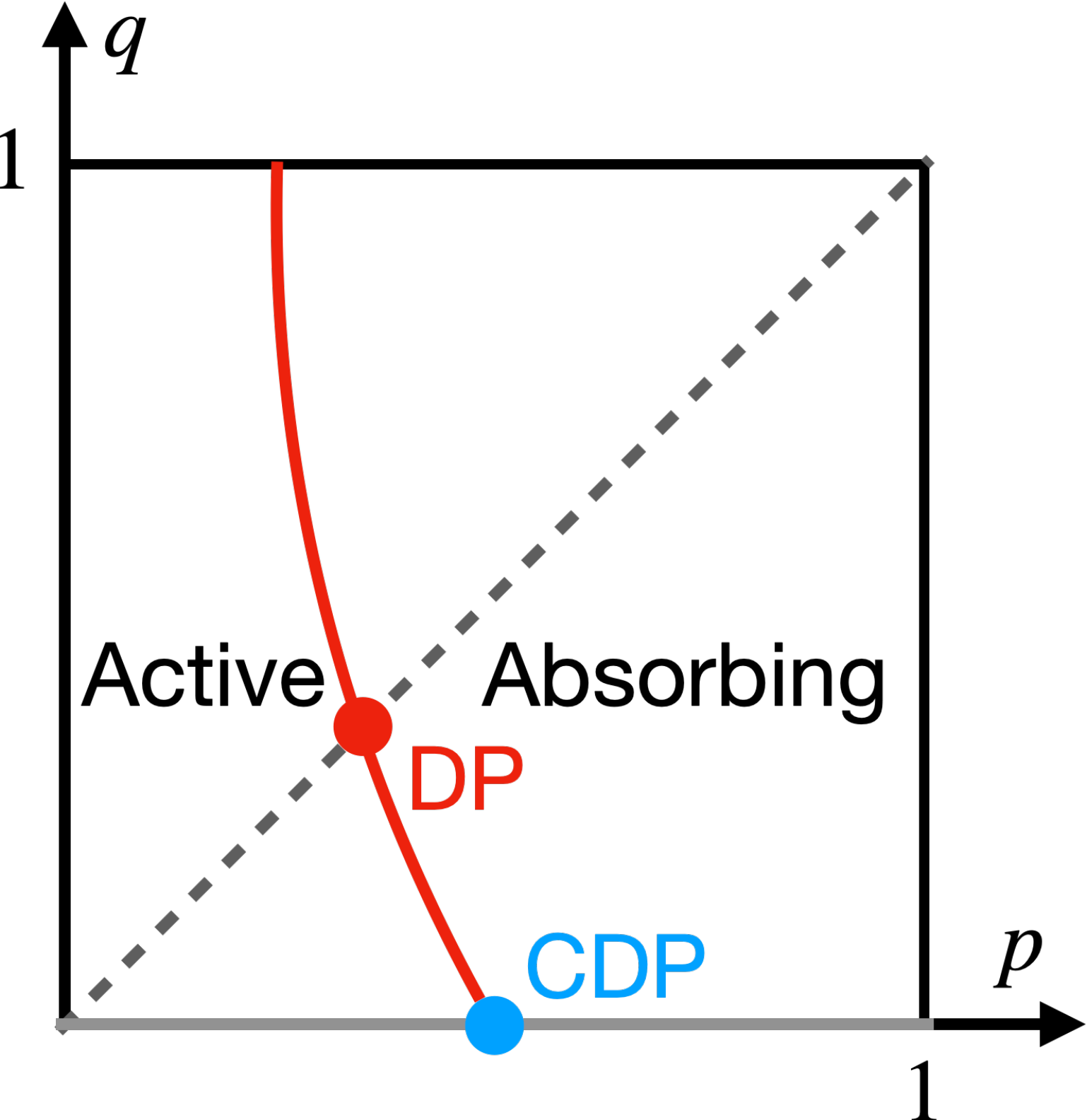}
\caption{ Schematic phase diagram of the Domany--Kinzel model. The critical behavior along the entire phase transition line (denoted as red in the figure) --- except for its lower terminal point --- is described by the DP universality class, while the critical behavior along the $q = 0$ line is described by the compact directed percolation (CDP). For the former, we focus on the diagonal line, $p = q$ (see main text).
}
\label{fig:DK_phase}
\end{figure}

The aforementioned developments have primarily been motivated from out-of-equilibrium quantum many-body systems. Since quantum statistical mechanics subsumes classical statistical mechanics, it is natural to ask whether these developments also shed new light on well-known classical non-equilibrium systems. Motivated from such considerations, we ask (i)
Do all steady states in the active regime of DP, corresponding to different values of the tuning parameter $\tau/g$, belong to the same phase of matter? (ii) How strongly non-Gibbsian is the DP critical point, as quantified by $\xi_{\mathrm{Markov}}$? 

Our main results are:

(a) We study a specific 1+1-D cellular automaton exhibiting phase transition in the DP universality class~\cite{domany1984equivalence}, and numerically demonstrate that the steady state corresponding to the critical point has a divergent Markov length, indicating that this state is strongly non-Gibbsian. It is interesting to contrast this with  equilibrium phase transitions and decoherence induced transitions in topological states -- in the former, the mutual information between far separated regions typically shows a singularity, and the CMI is non-singular, while in the latter, it is the other way around. In DP, both MI and CMI are singular at the critical point.
\\
\\
(b) Further, we provide numerical evidence that all steady states within the active phase are two-way connected via a locally reversible channel, providing support for an information-theoretic notion of phase equivalence.
\\
\\
(c) Along a special line in the phase diagram, the automaton we study exhibits a phase transition in the `compact directed percolation' (CDP) universality~\cite{domany1984equivalence, essam1989directed, janssen2005survival}. In this  case, the steady states are trivial but the dynamics remains nontrivial. For a certain initial state, we obtain an analytical expression for the CMI and demonstrate a diverging Markov length in a specific spatiotemporal limit. We relate this divergence to the spontaneous breaking of domain-wall parity symmetry from strong to weak.

\newpage

\textbf{Domany--Kinzel Model:} To understand (non)Markovianity and local reversibility for systems exhibiting DP, we will focus on a specific 1+1-D model introduced by Domany and Kinzel~\cite{domany1984equivalence}. The Domany–Kinzel (DK) model consists of a one-dimensional classical spin chain $\{ z_j \}_{j \in \mathbb{Z}}$, where each spin takes values $z_j = \pm 1$. We sometimes also denote $z_j = \uparrow (\downarrow)$ when $z_j = +1 (-1)$. At even (odd) times, only the odd (even) sites are updated according to the conditional probabilities $\mathcal{E}(z_{j,t+1} | z_{j-1,t}, z_{j+1,t})$, where $\mathcal{E}(\downarrow | \downarrow, \downarrow) = 1, \mathcal{E}(\downarrow | \uparrow, \uparrow) = q, 
\mathcal{E}(\downarrow | \uparrow, \downarrow) = \mathcal{E}(\downarrow | \downarrow, \uparrow) = p$. Note that these rules fully determine $\mathcal{E}$ due to probability conservation $\sum_{z'_j} \mathcal{E}(z'_j | z_{j-1}, z_{j+1}) = 1$. The schematic phase diagram of the Domany--Kinzel model is shown in Fig.~\ref{fig:DK_phase}. In the absorbing phase, the state $\rho_\downarrow = (|\downarrow\rangle \langle \downarrow|)^{\infty}$ is the unique steady state, while in the active phase, an additional steady state emerges, resulting in a two dimensional steady-state subspace. The critical behavior along the entire phase transition line (shown in red), except at its lower terminal point, is governed by the DP universality class. In contrast, the critical behavior along the $q = 0$ line belongs to the compact directed percolation (CDP) universality class~\cite{domany1984equivalence, essam1989directed, janssen2005survival}.

\begin{figure*}
\centering
\includegraphics[width=\linewidth]{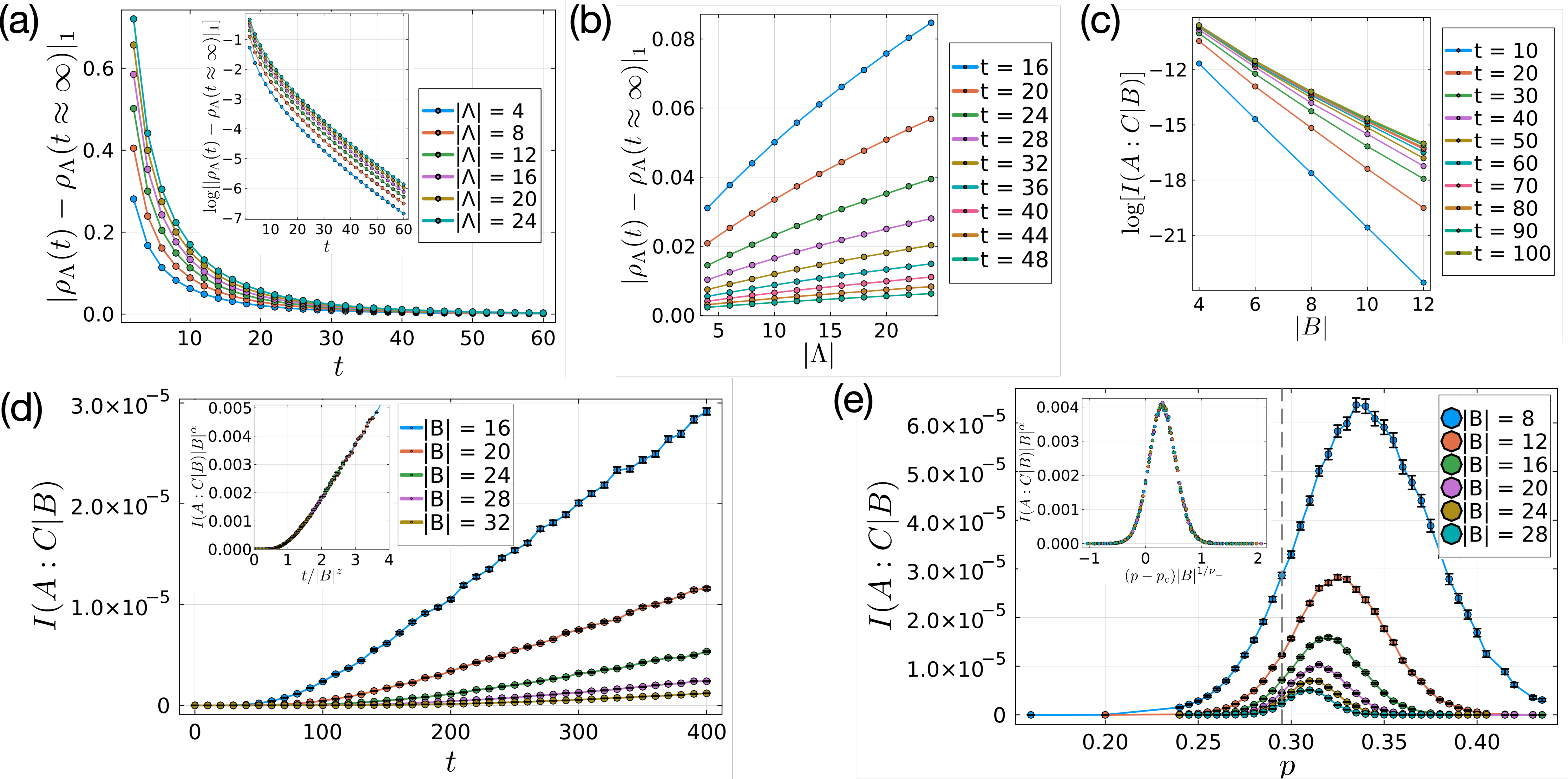}
\caption{
Local reversibility ((a), (b), (c)) and divergent Markov length ((d), (e)) in the Domany--Kinzel model along the $p=q$ line.
(a) $|\rho_\Lambda(t) - \rho_\Lambda(t \approx \infty)|_1 $ as a function of time $t$ with various $|\Lambda|$ at a representative $p < p_c$ where we choose $p = 0.24$.  
Here $t \approx \infty$ denotes the longest time (which is $310$ here) needed beyond which the magnetization does not change within the tolerance $10^{-10}$.
The inset shows the $\log |\rho_\Lambda(t) - \rho_\Lambda(t\approx \infty)|_1 $-$t$ plot. 
(b) $|\rho_\Lambda(t) - \rho_\Lambda(\infty)|_1 $ as a function of $|\Lambda|$ with various $t$.
(c) $\log I(A:C|B)$ as a function of $|B|$ for different times $t$, again at $p =0.24 $. The slope determines the Markov length $\xi$. We find that $\xi$ increases from $\xi \approx 0.67$ at $t = 10$ and saturates to $\xi \approx  1.27$ for $t \gtrsim 50$. 
(d) The conditional mutual information $I(A:C|B)$ as a function of time at $p = p_c \approx 0.295$. The inset shows the data collapse with the scaling ansatz Eq.\eqref{Eq:I_collpase_z}, with $z = 1.58$, $\alpha = 2$. (e) $I(A:C|B)$ as a function of the error rate $p$ for different size $|B|$ (the dashed line indicates $p_c$). The inset shows the data collapse of the scaling ansatz in Eq.\eqref{Eq:I_collpase_nu} with $\nu_{\perp} = 1.1$ and $\alpha = 2$. Here we fix $t/|B|^z = 2$ with $z = 1.58$. 
} 
\label{fig:CMI_stavskaya_combine}
\end{figure*}
\textbf{CMI along $p = q$ line (Directed Percolation):}
To access the DP transition, we study the DK model along the diagonal ($p = q$), which is equivalent to the Stavskaya's model~\cite{stavskaja1968homogeneous} and is expected to be representative of the behavior throughout the phase diagram away from $q = 0$ line.
Before we discuss our main results, let us consider a heuristic, mean-field argument that hints at the non-Markovianity of the steady state. Recall that the coarse-grained description is given by the Langevin dynamics $\partial_t{\phi} = \nabla^2 \phi + \tau \phi - g \phi^2/2 + \zeta$, with $\langle \zeta \rangle = 0$, and $\langle \zeta(\vec{x},t) \zeta(\vec{x'},t') \rangle = g \phi(x,t) \delta(\vec{x} - \vec{x'}) \delta(t - t')$.  At the mean-field level, in the active phase, $\phi = 2 \tau/g$. Let us improve the mean-field argument by taking into account Gaussian fluctuations.  We write $\phi(x,t) = \phi_0(t) + \delta \phi(x,t)$. The Fokker-Planck equation for the zero-momentum mode $\phi_0(t)$ can be solved exactly~\cite{janssen1988finite, lubeck2005finite}: $P_0(\phi_0) \propto e^{- V(\phi^2_0/2 - 2 \phi_0 \tau/g)}/{\phi_0}$, where $V = L^d$ is the total spatial volume of the system \footnote{Note that this probability distribution is not normalizable --- one can introduce an infinitesimal magnetic field to make it normalizable, see Ref.~\cite{lubeck2005finite}}. Within Gaussian approximation for the fluctuations $\delta \phi$, the effective noise for $\delta \phi$ satisfies $\langle \zeta(\vec{x},t) \zeta(\vec{x'},t') \rangle = g \phi_0(x,t) \delta(\vec{x} - \vec{x'}) \delta(t - t')$, i.e., the zero mode acts as an effective temperature for the Gaussian fluctuations (see Ref.\cite{supplement} for details). The steady state distribution for $\phi(x)$ within this approximation is therefore given by $P(\phi(x)) = P_0(\phi_0) e^{- \int_x \left(|\nabla \delta \phi(x)|^2 + \tau |\delta \phi(x)|^2\right)/(g\phi_0)}$, where, by definition, $\phi_0 = \int_x \phi(x)/L^d$ and $\delta \phi(x) = \phi(x) - \phi_0$. Due to the denominator $\phi_0$ in the exponential,  $-\log P(\phi(x))$ cannot be written as the integral of a local Hamiltonian, and thus the $P(\phi(x))$ is non-Markovian. A similar calculation for an equilibrium transition (e.g., the $O(N)$ model) leads to a Gibbs distribution with local Hamiltonian.

We now discuss our numerical results for CMI that have been obtained using tensor network simulations. Throughout, we will use the short-hand $|\rho\rangle = \sum_{\{z_j\}_{j \in Z}}\rho_{\{ z_j\}}|\{ z_j\}\rangle $ for the density matrix $\rho = \sum_{\{z_j\}_{j \in Z}}\rho_{\{ z_j\}}|\{ z_j\}\rangle \langle \{ z_j\}|$. The state vector $|\rho\rangle$ can be expressed as an matrix product state (MPS)\cite{blythe2007nonequilibrium,temme2010stochastic}, and the mixed state $|\rho(t)\rangle = \mathcal{E}^t|\rho_{\uparrow}\rangle$ as a function of time can then be studied using the infinite time-evolving block decimation (iTEBD) algorithm \cite{vidal2007classical}. {Our code implementation is publicly available at Ref.\footnote{See \url{https://github.com/yuhsuehchen/directed-percolation-cmi} for the code implementation in Julia.}}.
Throughout the simulation, the bond dimension of the MPS is set to $\chi = 48$. We verified that for $t \leq 400$—which, as we will show below, is sufficient to probe the critical phenomena—the quantitative behavior remains unchanged upon further increasing the bond dimension.  We choose the initial state as $\rho_{\uparrow} = (|\uparrow\rangle \langle \uparrow|)^{\infty}$, which is the exact long-time steady state when $p = q = 0$. If (i) $|\rho(t) - \rho(t=\infty)|_1 < \textrm{poly}(L) e^{-t/\tau}$ where $\tau$ is some $L$ independent constant, and (ii) the Markov length remains finite for all $t \lesssim \log(L)$, then that would imply that the steady state is in the same phase of matter as $\rho_{\uparrow}$~\cite{sang2024mixed,sang2025mixed}. 
The first condition is called rapid mixing \cite{aldous2006random, cubitt2015stability} and previously, has only been shown to hold true in an open ball around $p = q = 0$ \cite{de2012exponential}, while we are not aware of any results for the second condition. Now we will provide numerical evidence for both of these conditions for $p < p_c$. 

Since the system size we are simulating is infinite, the natural condition to verify is $|\rho_\Lambda(t) - \rho_\Lambda(\infty)|_1 < \mathrm{poly}(|\Lambda|) e^{-t/\tau}$ , where $\rho_\Lambda(t) \equiv \mathrm{Tr}_{\bar{\Lambda}}\rho(t)$, and $\Lambda$ denotes a connected, finite region ~\cite{cubitt2015stability}.  
Fig.\ref{fig:CMI_stavskaya_combine}(a) shows $|\rho_\Lambda(t) - \rho_\Lambda(t \approx \infty)|_1 $ as a function of time $t$ with various $|\Lambda|$ at a representative $p < p_c$ where we chose $p = 0.24$. Here $t \approx \infty$ denotes the longest time (which is $310$ at $p =0.24$) needed beyond which the magnetization does not change within the tolerance $10^{-10}$. 
We find that $|\rho_\Lambda(t) - \rho_\Lambda(t \approx \infty)|_1 $ decays exponentially with $t$ as shown in the inset. We further verify that for any fixed $t$, $|\rho_\Lambda(t) - \rho_\Lambda(\infty)|_1 $ grows at most polynomially in $|\Lambda|$, as shown in Fig.\ref{fig:CMI_stavskaya_combine}(b) (note that the rate of increase with $|\Lambda|$ is smaller at larger $t$). Finally, we remark that the rapid-mixing condition already implies that all observables $O$ with support $\Lambda$ satisfy $|O(t)-  O(t\approx \infty)|_1 <  \mathrm{poly}(|\Lambda|) e^{-t/\tau}$, see, e.g., Ref.\cite{supplement} where we show such data for magnetization.

Next, we consider CMI as a function of time and $|B|$, which is computed using the technique developed in Ref.~\cite{ferris2012perfect,lloyd2025diverging}. For numerical convenience, we choose both $A$ and $C$ to be single sites separated by a region $B$ of size $|B|$. We  also verified that increasing the size of region $C$ up to $|C| = 10$ sites does not change the results qualitatively. Fig.~\ref{fig:CMI_stavskaya_combine}(c) plots the logarithm of the CMI as a function of $|B|$ with different $t$, again at $p = 0.24$. The slope determines the Markov length $\xi_{\text{Markov}}$. We find that $\xi_{\text{Markov}}$ gradually increases from approximately $0.67$ at $t = 10$ and saturates to $\xi_{\text{Markov}} \approx 1.27$ for $t \gtrsim 50$. In the steady state, $I(t \approx \infty) \sim e^{-|B|/\xi_{\textrm{Markov}}}$ as expected, and crucially, the Markov length remains finite throughout the evolution.

Having established that the Markov length $\xi_{\text{Markov}}$ remains finite when $p < p_c$, we now present numerical evidence that it diverges at the critical point. First, we verify that the system does not satisfy the aforementioned rapid mixing property, as indicated by the polynomial convergence of magnetization with time~\cite{supplement,mendoncca2011monte}. Fig.~\ref{fig:CMI_stavskaya_combine}(d) shows the CMI at $p = p_c \approx 0.295$ as a function of time for different values of $|B|$. We find that, for any fixed $|B|$, the CMI increases monotonically with time up to $t = 400$, which is the maximum time accessible in our numerics. Conversely, for any fixed $t$, the CMI decreases as $|B|$ increases. Notably, the data for different system sizes collapse almost perfectly under the scaling ansatz
\begin{equation}
\label{Eq:I_collpase_z}
I(A:C|B) = \frac{1}{|B|^{\alpha}} f_1\Big( \frac{t}{|B|^z} \Big),
\end{equation}
with the choices $z = 1.58$ and $\alpha = 2$ [see the inset of Fig.~\ref{fig:CMI_stavskaya_combine}(d)]. We note that the best fit for $z$ matches rather well with the dynamical exponent for $(1\!+\!1)$-D DP, $z = 1.5807$~\cite{hinrichsen2000non}. Assuming that, for fixed $|B|$, the CMI saturates to a non-zero constant in the limit $t \rightarrow \infty$ (i.e., $f_1(\infty) = \mathrm{const}$), we conclude that the long-time CMI at the critical point decays polynomially as a function of $|B|$. Therefore, the Markov length $\xi_{\text{Markov}}$ (recall $I(A:C|B) \sim e^{-|B|/\xi_{\text{Markov}}}$) diverges at the critical point. We observe that exponent $\alpha \approx 2$ for CMI is quite different than that for the two-point correlations of magnetization, $\langle m(x) m(0) \rangle \approx 1/x^{0.504}$~\cite{hinrichsen2000non}, which lower bounds the decay of the mutual information between $A$ and $B$, $I(A:B) \geq 1/{r_{AB}}^{1.008}$.

To further identify the Markov-length exponent defined through $\xi_{\text{Markov}} \sim |p - p_c|^{\nu}$ and to provide stronger evidence that $\xi_{\text{Markov}}$ remains finite when $p < p_c$, Fig.~\ref{fig:CMI_stavskaya_combine}(e) shows the CMI as a function of $p$ for different system sizes, with the time fixed at $t = 2|B|^{z}$. We have also examined the cases $t/|B|^{z} = 1$ and $3$, and the resulting curves are similar. In Fig.~\ref{fig:CMI_stavskaya_combine}(e), one clearly observes that the peak of the CMI approaches $p_c \approx 0.295$ as the system size increases.  Furthermore, the data from different system sizes collapse almost perfectly under the scaling ansatz
\begin{equation}
\label{Eq:I_collpase_nu}
I(A:C|B) = \frac{1}{|B|^{\alpha}} f_2\Big( (p - p_c) |B|^{1/\nu} \Big),
\end{equation}
with $\alpha = 2$ (the same value used in Eq.~\eqref{Eq:I_collpase_z}) and $\nu = 1.1$ [see the inset of Fig.~\ref{fig:CMI_stavskaya_combine}(e)]. We note that the fitted value of $\nu$ matches the perpendicular correlation-length exponent $\nu_{\perp} = 1.0969$~\cite{hinrichsen2000non}. This establishes that the Markov length diverges with the same critical exponent as the standard correlation length.

\textbf{CMI along $q = 0$ line (Compact Directed Percolation):} 
To build analytical intuition for non-Markovianity in the DK model, we now consider the CMI along the $q =0$ line.  Along this special line, unlike the standard DP, the steady-state subspace remains two-dimensional for all $p$. To see this explicitly, we rewrite the aforementioned update rules at $q = 0$ as $\mathcal{E}(\downarrow | \downarrow, \downarrow) = \mathcal{E}(\uparrow | \uparrow, \uparrow) = 1, \mathcal{E}(\downarrow | \uparrow, \downarrow) = \mathcal{E}(\downarrow | \downarrow, \uparrow) = p$. This implies that, apart from $\rho_{\downarrow} = (|\downarrow\rangle \langle \downarrow|)^{\infty}$, $\rho_{\uparrow} = (|\uparrow\rangle \langle \uparrow|)^{\infty}$ is also an absorbing state. 
Therefore, the steady-state subspace is identical for all $p$, and thus the question about the two-way connection between steady states at different $p$ is not interesting. 
Despite this, one can still observe a `phase transition' across $p = p_c = 1/2$ if one studies  dynamics of a generic initial state~\cite{domany1984equivalence, essam1989directed, janssen2005survival}.  Remarkably, as we demonstrate now, along this line, for a specific initial state, one can obtain an analytical expression for the time-dependent CMI with a clear physical interpretation. We will find that in a certain spatiotemporal limit, the Markov length diverges for all $p$, but the CMI can still detect the aforementioned phase transition.

We consider the initial state $|\rho_0\rangle = |\rho_{\text{dw, sym}}\rangle \equiv \frac{1}{2} |\rho_{\text{dw}, \uparrow} \rangle + \frac{1}{2} |\rho_{\text{dw}, \downarrow} \rangle$ where $|\rho_{\text{dw}, \uparrow} \rangle = |\cdots, \uparrow_{-5}, \uparrow_{-3}, \uparrow_{-1}, \downarrow_{1}, \downarrow_{3}, \downarrow_{5}, \cdots \rangle$, and $|\rho_{\text{dw}, \downarrow}\rangle = \prod_j X_j |\rho_{\text{dw}, \uparrow}\rangle$ is its $\mathbb{Z}_2$-symmetric partner. Recall that at even times, only the odd sites are updated; therefore, it suffices to specify the configuration of the odd-indexed sites at $t = 0$. 
To compute CMI, we divide the system into three regions, $A$, $B$, and $C$, by choosing $A = [-\infty, \cdots, -2r]$, $B = [-2r, \cdots, 2r-1]$, and $C = [2r, \cdots, \infty]$, which correspond to the left, middle, and right subsystems, respectively. A key point is that the evolution can be mapped to a directed random walk for the domain wall present in the initial state~\cite{domany1984equivalence}. We find that, as a function of time $t = 2\tau$, the CMI $I(A:C|B)$ for the evolved state $|\rho(t)\rangle = \mathcal{E}^t |\rho_0\rangle$ is given by~\cite{supplement}:
\begin{equation}
\label{Eq:CMI_pi}
I(A:C|B) = \big(\sum_{b \notin S_{B,\text{dw}}} \Pi(B = b)\big) \log 2,
\end{equation}
where $\Pi(B = b)$ denotes the marginal probability of the configuration $B =b$ and $S_{B,\text{dw}}$ denotes all configurations in which the domain wall lies within region $B$. Eq.\eqref{Eq:CMI_pi} thus has a clear physical interpretation: when domain wall lives in region $B$, then all correlations between $A$ and $C$ are mediated via $B$, and CMI is proportional to the probability that the domain wall escapes  region $B$. %

Using Eq.\eqref{Eq:CMI_pi}, one can show that the CMI vanishes for $\tau \leq r$. When $\tau > r$, in the large-$\tau$, large-$r$ limit, one finds the following asymptotic form \cite{supplement}:
\begin{equation}
\label{Eq:CMI_asymp}
I(A:C|B) = \frac{\log 2}{2} \Big[ \text{erfc}\!\Big( \frac{1 - 2 u v}{\sqrt{v}} \Big) + \text{erfc}\!\Big( \frac{1 + 2 u v}{\sqrt{v}} \Big) \Big],
\end{equation}
where $u = (p - p_c) r$ with $p_c = 1/2$, $v = \tau / r^2$, and $\text{erfc}(x) = \frac{2}{\sqrt{\pi}} \int_x^{\infty} e^{-s^2} \, ds$ is the complementary error function. Eq.~\eqref{Eq:CMI_asymp} implies the scaling exponents $z = 2$ and $\nu_\perp = 1$, consistent with those of the $(1+1)$-dimensional CDP universality.  Another interesting feature of this choice of initial state is that
\begin{equation}
\begin{aligned}
\lim_{r \rightarrow \infty} \lim_{\tau \rightarrow \infty} I(A:C|B) = \log 2,
\end{aligned}
\end{equation}
\textit{independent} of $p$, indicating that nonlocal information is encoded in the system -- this is again related to the fact that in this limit, the domain wall always escapes region $B$, thereby generating non-zero CMI.  Therefore, for this initial state, the Markov length always diverges in this spatiotemporal limit. We emphasize that the order of limits is crucial: reversing the order gives a trivial one since $I(A:C|B) = 0$ when $\tau \leq r$. The divergence of the Markov length throughout the whole phase is reminiscent of the strong-to-weak symmetry breaking~\cite{lee2023quantum,lessa2024strong}. In fact, the long-time state can be shown to spontaneously break the symmetry associated with the conservation of the domain-wall parity from strong to weak~\cite{supplement}.

We anticipate that any generic initial state should be able to probe the transition. We also considered a more natural choice of initial state which has only one active site localized at the origin and study the CMI numerically~\cite{supplement}. We find that the CMI yields the same exponents $z$ and $\nu_{\perp}$. However, with this initial condition, instead of saturating to $\log 2$ in the large-$\tau$ limit, the CMI away from $p = p_c$ at a fixed $v = \tau/r^2$ vanishes exponentially fast in $r$. In contrast,  at $p=p_c$, it exhibits a power-law behavior as a function of $r$ at any fixed $v$.

Before closing, we note a few promising directions. It would be interesting to also investigate other classical non-equilibrium states from the lens of local reversibility and Markov length, such as those that arise in the context of surface growth models~\cite{kardar1986dynamic}, flocking~\cite{toner1995long}, or classical memory without symmetry~\cite{toom1974nonergodic, toom1980stable}. One specific question is: consider a non-equilibrium system where terms that break detailed balance are `dangerously irrelevant' in the sense that they are irrelevant at the critical point in the RG sense, but are essential for the physics of the proximate phase (an example is provided by the Toom's PCA~\cite{toom1974nonergodic, toom1980stable,he1990generic}). Does the Markov length remains finite and non-singular across such transitions? Given the central role played by conditional mutual information in our discussion, another worthwhile direction is constraints on the RG flow in non-equilibrium from information-theoretic constraints such as strong subadditivity.

\textbf{Acknowledgments:}
We thank Shengqi Sang for useful discussions, and John McGreevy, Sarang Gopalakrishnan and Tim Hsieh for  feedback on the manuscript. T.G. is supported by
the National Science Foundation under Grant No. DMR-2521369. We acknowledge the hospitality of
Kavli Institute for Theoretical Physics (KITP) and thank
the organizers of the KITP programs ``Noise-robust Phases of Quantum Matter'', and ``Learning the Fine Structure of Quantum Dynamics in Programmable Quantum Matter''. This research was supported in part by grant NSF PHY-2309135
to the KITP.

%

\onecolumngrid 

\clearpage %

\appendix

\section{Steady-state probability distribution for DP within mean-field supplemented with fluctuations}

{The goal of this appendix is to provide a heuristic derivation of the non-Markovianity of the steady state of directed percolation in general dimensions by taking into account the Gaussian fluctuations on top of the mean-field arguments.} Recall that the Langevin equation for the coarse-grained field $\phi(x,t)$ is

\be 
\partial_t{\phi(x,t)} = \nabla^2 \phi(x,t) + \tau \phi(x,t) - \frac{g}{2} \phi(x,t)^2 + \zeta(x,t) \label{eq:langevin_full}
\ee 
with $\langle \zeta(x,t) \rangle = 0$ and $\langle \zeta(x,t) \zeta(x',t') \rangle = g \phi(x,t) \delta(x-x') \delta(t - t')$. We will focus on the active phase, which corresponds to $\tau > 0$. Within mean-field, the activity in this phase is $\phi = 2 \tau/g$.

We  decompose $\phi(x,t)$ as $\phi(x,t) = \phi_0(t) + \delta \phi(x,t)$, where $\phi_0(t) = \int_x \phi(x,t)/V$ is the zero momentum mode, and $\delta \phi(x,t)$ denotes fluctuations around the zero mode ($V = L^d$ is the total volume of the system). To obtain a Fokker-Planck (FP) equation for the zero mode $\phi_0$, we perform a spatial average of the Langevin equation. Dropping terms quadratic in $\delta \phi$, one obtains

\be 
\partial_t \phi_0(t) = \tau \phi_0(t) - \frac{g}{2} \phi_0(t)^2 + \int_x \frac{\zeta(x,t)}{V} \label{eq:langevin_zeromode}
\ee 

It is convenient to define rescaled variables, $\phi'_0(t) = \phi_0 \sqrt{V}$, $s = gt/\sqrt{V}$, $T = \tau \sqrt{V}/g$, so that the Langevin equation can be written as,

\be 
\partial_s \phi'_0(s) = T \phi'_0(s) - \frac{1}{2}{\phi'_0}^2(s) + \zeta'(s),
\ee 
where $\langle \zeta'(s) \zeta'(s') \rangle =  \phi'_0(s)\delta(s-s')$ and we have again kept terms only to the leading non-zero order. This leads to the following Fokker-Planck equation for the probability distribution of $\phi'_0$:

\be 
\partial_s P(\phi'_0, s) = \left[ \partial_{\phi'_0} \left(-T \phi'_0 + \frac{{\phi'_0}^2}{2}\right) + \partial_{\phi'_0}^2 \frac{\phi'_0}{2}\right] P(\phi'_0,s)
\ee 
This equation is identical to the one derived in Refs.~\cite{janssen1988finite, lubeck2005finite} using a path integral approach, and the steady-state solution is:
\be 
P(\phi'_0) \propto \frac{e^{- (-2T + \phi'_0/2)  \phi'_0}}{\phi'_0} \quad \Rightarrow \quad P(\phi_0) \propto \frac{e^{- V(\phi^2_0/2 - 2 \phi_0 \tau/g)}}{\phi_0}
\ee 
Therefore, the probability distribution is sharply peaked around $\phi_0 = 2 \tau/g$, as expected. Note that this distribution is not normalizable and to regularize it, one needs to add an infinitesimal source field $h$ to the Langevin equation (Eq.\ref{eq:langevin_full}). In the presence of this field, the solution is~\cite{lubeck2005finite}
$P(\phi'_0) = {\phi'_0}^{2 H-1} e^{- (2T + \phi'_0/2) \phi'_0}$ where $H = V h/g$.

Now let us find the steady state probability distribution for the fluctuations $\delta \phi(x)$ around $\phi_0$. The corresponding Langevin equation is:

\be 
\partial_t \delta \phi(x,t) = \nabla^2 \delta \phi(x,t) - \tau \delta \phi(x,t)+ \zeta(x,t)
\ee 
where we have employed the fact that $\phi_0$ is sharply peaked at $2 \tau/g$, and we have dropped terms non-linear in $\delta \phi$. Crucially, the noise-noise correlator is $\langle \zeta(x,t) \zeta(x,t) \rangle = g \phi(x,t)   \delta(x-x') \delta(t - t') \approx g \phi_0(t)   \delta(x-x') \delta(t - t')$, and therefore, the zero mode $\phi_0(t)$ acts as an effective temperature for the fluctuations. The steady state distribution within this approximation for $\delta \phi(x)$, at a fixed value of $\phi_0(x)$,  is therefore given by $P(\delta \phi(x)|\phi_0) = e^{- \int_x \left(|\nabla \delta \phi(x)|^2 + \tau |\delta \phi(x)|^2\right)/\left(g\phi_0\right)}$. The steady-state distribution for the full field $\phi(x)$ is therefore,

\be 
P(\phi(x)) = P(\phi_0)\, P(\delta \phi(x)|\phi_0(x))
\ee 

Due to the denominator $\phi_0$ in the exponential,  $-\log P(\phi(x))$ cannot be written as the integral of local Hamiltonian, and thus the $P(\phi(x))$ is very non-Gibbsian.

\section{Derivation of CMI for Compact Directed Percolation}
The goal of this appendix is to provide detailed derivation of the CMI for compact directed percolation (CDP), which corresponds to the $q = 0$ line in the main text. Along the $q = 0$ line, at even (odd) times, the odd (even) sites are updated according to the conditional probabilities $\mathcal{E}(z_{j, t+1}|z_{j-1,t}, z_{j+1,t})$, where
\begin{equation}
\label{Eq:supp_dk_PCA_q0}
\begin{aligned}
\mathcal{E}(\downarrow | \downarrow, \downarrow) & = \mathcal{E}(\uparrow | \uparrow, \uparrow) = 1, \\
\mathcal{E}(\downarrow | \uparrow, \downarrow) & = \mathcal{E}(\downarrow | \downarrow, \uparrow) = p.
\end{aligned}
\end{equation}
We will first consider the case discussed in the main text, where the initial state has odd number of domain walls. In particular, we will derive the analytical form of Eq.\eqref{Eq:CMI_pi} and Eq.\eqref{Eq:CMI_asymp} in the main text. We will later study the CMI of the case where the initial state has an even number of domain walls numerically. 

\subsection{Odd Number of Domain Walls}

\subsubsection*{Conditional Mutual Information}

\begin{figure*}
\centering
\includegraphics[width=\linewidth]{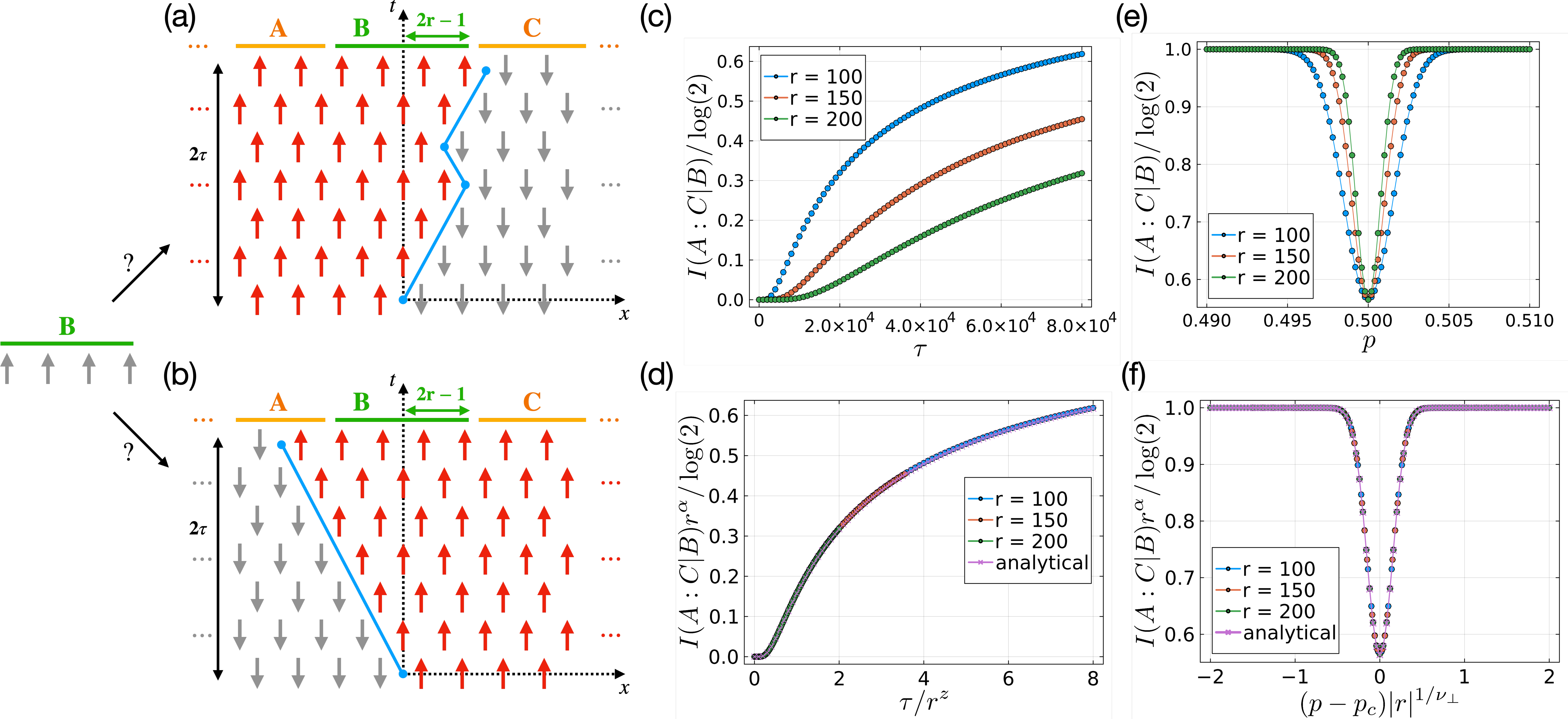}
\caption{
Conditional mutual information (CMI) analysis of the compact directed percolation model with the initial state 
$|\rho_{\text{dw, sym}}\rangle \equiv \tfrac{1}{2} |\rho_{\text{dw}, \uparrow}\rangle + \tfrac{1}{2} |\rho_{\text{dw}, \downarrow}\rangle$, 
where one representative trajectory of $|\rho_{\text{dw}, \uparrow}\rangle$ and $|\rho_{\text{dw}, \downarrow}\rangle$ is shown in (a) and (b), respectively 
[see also Eq.~\eqref{Eq:suppp_rho0_CDP}]. 
If $b = |\uparrow\rangle^{|B|}$, there is an ambiguity in inferring whether the domain wall lies in region $C$ [as shown in (a)] or in region $A$ [as shown in (b)]. 
(c) $I(A:C|B)$ as a function of time at $p = p_c = 0.5$. 
(d) Finite-size, finite-time scaling collapse with $z = 2$ and $\alpha = 0$, compared with the analytical result derived in Eq.~\eqref{Eq:supp_CMI_asymp}. 
(e) $I(A:C|B)$ as a function of the error rate $p$ for different sizes $r$, where we fix $\tau/r^z = 6$ with $z = 2$. 
(f) Finite-size, finite-time scaling collapse with $\nu_{\perp} = 1$ and $\alpha = 0$, compared with the analytical result derived in Eq.~\eqref{Eq:supp_CMI_asymp}.
}
\label{fig:CDP_full}
\end{figure*}

Let's first recall the setup in the main text.  We consider the initial state at time $t=0$
\begin{equation}
\label{Eq:suppp_rho0_CDP}
|\rho_0\rangle = |\rho_{\text{dw, sym}}\rangle \equiv \frac{1}{2} |\rho_{\text{dw}, \uparrow} \rangle + \frac{1}{2} |\rho_{\text{dw}, \downarrow} \rangle,
\end{equation}
where  
\begin{equation}
\label{Eq:rho_dw_up}
|\rho_{\text{dw}, \uparrow} \rangle = |\cdots, \uparrow_{-5}, \uparrow_{-3}, \uparrow_{-1}, \downarrow_{1}, \downarrow_{3}, \downarrow_{5}, \cdots \rangle
\end{equation}
[see Fig.~\ref{fig:CDP_full}(a)], and $|\rho_{\text{dw}, \downarrow}\rangle = \prod_j X_j |\rho_{\text{dw}, \uparrow}\rangle$ [see Fig.~\ref{fig:CDP_full}(b)] is its $\mathbb{Z}_2$-symmetric partner. Note that we do not specify the configuration on even sites in the initial condition since the evolution is insensitive to it (because at each time step, odd sites dictate the spin configuration for even sites at the next time step and vice-versa). Therefore, with a slight abuse of notation, we will only label the spin configuration on odd sites at even times, and even sites at odd times.
We divide the system into three regions, $A$, $B$, and $C$, by choosing $A = [-\infty, \cdots, -2r]$, $B = [-2r+1, \cdots, 2r-1]$, and $C = [2r, \cdots, \infty]$, which correspond to the left, middle, and right subsystems, respectively [see Fig.~\ref{fig:CDP_full}(a) for an example with $r = 2$]. We are interested in the CMI $I(A:C|B)$ for the evolved state $|\rho(t)\rangle = \mathcal{E}^t |\rho_0\rangle$ as a function of $t = 2\tau$ and $|B | = 4r-2$.

To derive Eq.\eqref{Eq:CMI_pi} and Eq.\eqref{Eq:CMI_asymp}, we begin by considering the motion of the ``domain wall'' in Eq.\eqref{Eq:rho_dw_up}. A domain wall on site $j$ is defined as $\sigma_j \equiv z_{j-1} z_{j+1} = -1$ ($\sigma_j = +1$ indicates the absence of a domain wall). There are two types of domain walls, $\uparrow\downarrow$ and $\downarrow\uparrow$, and this distinction will be crucial when computing the CMI. The state $|\rho_{\text{dw}, \uparrow}\rangle$ in Eq.~\eqref{Eq:rho_dw_up} therefore corresponds to a single domain wall located at the origin, with $\sigma_0 = z_{-1} z_1 = -1$.  The first line of Eq.~\eqref{Eq:supp_dk_PCA_q0} enforces the constraint that a domain wall cannot be created from scratch. In contrast, the second line of Eq.~\eqref{Eq:supp_dk_PCA_q0} implies that a domain wall moves toward the all-$\uparrow$ (all-$\downarrow$) direction with probability $p$ ($1-p$) until it encounters another domain wall, at which point the pair annihilates. For instance, when $p = 1$, the domain wall in Eq.~\eqref{Eq:rho_dw_up} deterministically shrinks to the left with a constant velocity $v = 1$. On the other hand, for $p \in (0,1)$, the domain wall can move either left or right with probabilities $p$ and $(1-p)$, respectively. Fig.~\ref{fig:CDP_full}(a) illustrates a particular trajectory after time $t = 6$: since the domain wall moves once toward the $\uparrow$ direction and five times toward the $\downarrow$ direction, the probability of this trajectory is $p(1-p)^5$.

Given the above intuitive understanding of the domain-wall motion in $|\rho_{\text{dw}, \uparrow}\rangle$, one may wonder whether the evolved state $|\rho_{\text{dw}, \uparrow}(t)\rangle = \mathcal{E}^t |\rho_{\text{dw}, \uparrow}\rangle$ is sufficient to capture the transition without considering the symmetric state $|\rho_{\text{dw, sym}}\rangle$ defined in Eq.~\eqref{Eq:suppp_rho0_CDP}. However, we now show that the CMI for $|\rho_{\text{dw}, \uparrow}(t)\rangle$ is identically zero, independent of the time $t$ and the width $|B|$.

Recall that the CMI can be expressed as the average mutual information between $A$ and $C$ conditioned on $B$. Denoting by $S_X$ the set of all possible spin configurations in region $X$ and by $\Pi(x)$ the corresponding marginal probability, we can write
\begin{equation}
\label{Eq:supp_cmi_condB}
\begin{aligned}
I(A:C|B) & = \sum_{b \in S_B} \Pi(B = b)\, I(A:C|B = b),
\end{aligned}
\end{equation}
where
\begin{equation}
I(A:C|B = b) = \sum_{a \in S_A,\, c \in S_C} \Pi(a, c | b) 
\log \!\Bigg[ \frac{\Pi(a, c | b)}{\Pi(a | b)\, \Pi(c | b)} \Bigg]
\end{equation}
is the mutual information between $A$ and $C$ conditioned on $B = b$.
Note that if either $\Pi(a|b)$ or $\Pi(c|b)$ is a delta function (i.e., the configuration in $A$ or $C$ is deterministic given $B$), then $I(A:C|B = b) = 0$. We now show that this is indeed the case for all $b \in S_B$ when choosing $|\rho_{\text{dw}, \uparrow}\rangle$ as the initial state.

Let us divide the set $S_B$ into two disjoint subsets, $S_{B,\text{dw}}$ and $S_B \setminus S_{B,\text{dw}}$, where $S_{B,\text{dw}}$ denotes all configurations in which the domain wall lies within region $B$. When $b \in S_{B,\text{dw}}$, both $a \in S_A$ and $c \in S_C$ are deterministic, taking the values $|\uparrow\rangle^{|A|}$ and $|\downarrow\rangle^{|C|}$, respectively. Therefore, $I(A:C|B = b) = 0$ for all $b \in S_{B,\text{dw}}$. On the other hand, if $b \notin S_{B,\text{dw}}$, then $b$ must be either $|\uparrow\rangle^{|B|}$ or $|\downarrow\rangle^{|B|}$. In this case, the configuration $b$ fully determines the spin configuration in either $A$ or $C$. For example, if $b = |\uparrow\rangle^{|B|}$, then $a \in S_A$ is fixed to be $|\uparrow\rangle^{|A|}$ [see Fig.~\ref{fig:CDP_full}(a)]. This follows from the fact that the domain wall in $|\rho_{\text{dw}, \uparrow}\rangle$ has a fixed orientation from $\uparrow$ to $\downarrow$, and the update rule in Eq.~\eqref{Eq:supp_dk_PCA_q0} does not change this orientation. Hence, the evolved state $|\rho_{\text{dw}, \uparrow}(t)\rangle = \mathcal{E}^t |\rho_{\text{dw}, \uparrow}\rangle$ always has vanishing CMI, independent of $t$, $|B|$, and $p$.

Given the above analysis of $|\rho_{\text{dw}, \uparrow}(t)\rangle$, we are now ready to derive Eq.~\eqref{Eq:CMI_asymp}, which gives the CMI for $|\rho(t)\rangle = \mathcal{E}^t |\rho_{\text{dw, sym}}\rangle$, where $|\rho_{\text{dw, sym}}\rangle = (|\rho_{\text{dw}, \uparrow}\rangle + |\rho_{\text{dw}, \downarrow}\rangle)/2$. A key advantage of choosing $|\rho_{\text{dw, sym}}\rangle$ as the initial state is that the time-evolved state $|\rho(t)\rangle$ respects the parity symmetry $x \rightarrow -x$ at all times, which greatly simplifies the calculation of the CMI. 

We now show that the CMI of $|\rho(t)\rangle$ can be written as
\begin{equation}
\label{Eq:supp_CMI_pi}
I(A:C|B) = \log 2 \Bigg( \sum_{b \notin S_{B,\text{dw}}} \Pi(B = b) \Bigg),
\end{equation}
where we have followed the same construction as before by dividing the set of all possible configurations $S_B$ into two disjoint subsets, $S_{B,\text{dw}}$ (where the domain wall lies in $B$) and $S_B \setminus S_{B,\text{dw}}$. 
We first note that if $b \in S_{B,\text{dw}}$, then $a \in S_A$ and $c \in S_C$ remain fully deterministic. This is because the single domain walls encoded in $|\rho_{\text{dw}, \uparrow}\rangle$ and $|\rho_{\text{dw}, \downarrow}\rangle$ are of opposite types (one $\uparrow\downarrow$, the other $\downarrow\uparrow$), and the update rule in Eq.~\eqref{Eq:supp_dk_PCA_q0} cannot change the domain-wall type. Consequently, one can always infer whether the trajectory originated from $|\rho_{\text{dw}, \uparrow}\rangle$ or $|\rho_{\text{dw}, \downarrow}\rangle$, thereby determining the spin configurations in $A$ and $C$.

On the other hand, if $b \notin S_{B,\text{dw}}$, the information contained in region $B$ is not sufficient to determine the origin of the trajectory. For example, $b = |\uparrow\rangle^{|B|}$ can originate either from $|\rho_{\text{dw}, \uparrow}\rangle$, with the domain wall moving to the right and ending in $C$ [Fig.~\ref{fig:CDP_full}(a)], or from $|\rho_{\text{dw}, \downarrow}\rangle$, with the domain wall moving to the left and ending in $A$ [Fig.~\ref{fig:CDP_full}(b)]. Therefore, it is natural to expect that an additional bit of information is encoded nonlocally in $A$ and $C$, such that 
\begin{equation}
\label{Eq:IAC_b}
I(A:C|b \notin S_{B,\text{dw}}) = \log 2,
\end{equation}
\textit{irrespective} of $t$, $|B|$, and $p$. To demonstrate Eq.~\eqref{Eq:IAC_b} more explicitly, note that if $b \notin S_{B,\text{dw}}$, the domain wall must lie in either $A$ or $C$. Without loss of generality, assume $a \in S_{A,\text{dw}}$ and thus the domain wall is of the $\downarrow\uparrow$ type [which also implies $b = |\uparrow\rangle^{|B|}$; see Fig.~\ref{fig:CDP_full}(b)]. 
Similar to the case where $b \in S_{B,\text{dw}}$, this implies $c = |\uparrow\rangle^{|C|}$ is deterministic, and thus one has $\Pi(a|b) = \Pi(a,c|b)$. On the other hand, one has $\Pi(c = |\uparrow\rangle^{|C|} | b = |\uparrow\rangle^{|B|}) = \sum_{a \in S_{A,\text{dw}}} \Pi(a | b = |\uparrow\rangle^{|B|})$ and $\Pi(c = |\uparrow\rangle^{|C|} | b = |\uparrow\rangle^{|B|}) = 1 - \sum_{c \in S_{C,\text{dw}}} \Pi(c | b = |\uparrow\rangle^{|B|})$ for any $c \notin S_{C,\text{dw}}$. Combining these with the inversion symmetry $\sum_{c \in S_{C,\text{dw}}} \Pi(c|b) = \sum_{a \in S_{A,\text{dw}}} \Pi(a|b)$, one obtains $\Pi(c|b) = 1/2$. Therefore, ${\Pi(a,c|b)}/{[\Pi(a|b)\Pi(c|b)]} = 2$, and thus Eq.~\eqref{Eq:IAC_b} follows.

Having established Eq.\eqref{Eq:supp_CMI_pi}, we now derive the analytical form of CMI. We first note that $\sum_{b \notin S_{B,\text{dw}}} \Pi{(B = b)} = \sum_{a \in S_{A,\text{dw}}} \Pi{(A = a)}+ \sum_{c \in S_{C,\text{dw}}} \Pi{(C = c)} = 2\sum_{c \in S_{C,\text{dw}}} \Pi{(C = c)}$ by the reflection symmetry.
Recall we choose $t = 2 \tau$, $A = [-\infty, \cdots, -2r]$, $B = [-2r+1,\cdots,2r-1]$, and $C=[2r, \cdots,\infty]$ belongs to the rest of the system [see Fig.\ref{fig:CDP_full}(a) for an example with $r = 2$].
Since $|\rho_{\text{dw, sym}}\rangle  = \frac{1}{2}|\rho_{\text{dw}, \uparrow }\rangle + \frac{1}{2}|\rho_{\text{dw} \downarrow}\rangle
$, the probability that the domain wall lies in $C$ can be contributed from both $|\rho_{\text{dw},\uparrow} \rangle$ and $|\rho_{\text{dw},\downarrow} \rangle$. 
Recall that at each time step, the domain wall has to move to either right or left.
For both of them to reach $C$, they have to move right with at least $R \geq \tau +r$ steps so that $L \leq \tau -r$ and thus $R-L \geq 2r$. The difference between $|\rho_{\text{dw},\uparrow} \rangle$ and $|\rho_{\text{dw},\uparrow} \rangle$ lies in the fact that the probability for $|\rho_{\text{dw},\uparrow} \rangle$ ($|\rho_{\text{dw},\uparrow} \rangle$) to move right in each step is $1-p (p)$. 
Therefore, the corresponding probabilities are $\sum_{R = \tau+r}^{2 \tau} \binom{2\tau}{R} (1-p)^{R} p^{2\tau-R} /2$ and $\sum_{R = \tau+r}^{2 \tau} \binom{2\tau}{R} p^{R} (1-p)^{2\tau-R} /2$ and for $|\rho_{\text{dw},\uparrow} \rangle$ and $|\rho_{\text{dw},\uparrow} \rangle$, respectively. It follows that
\begin{equation}
\begin{aligned}
 \sum_{b  \in S_{B,\text{dw}}} \Pi{(B = b)} = \sum_{R = \tau+r}^{2 \tau} \binom{2\tau}{R} (1-p)^{R} p^{2\tau-R} +  \sum_{R = \tau+r}^{2 \tau} \binom{2\tau}{R} p^{R} (1-p)^{2\tau-R}.
\end{aligned}
\end{equation}
In the large-$t$, large-$r$ limit:
\begin{equation}
\label{Eq:supp_CMI_asymp}
I(A:C|B) = \frac{\log 2}{2} \Big[ \text{erfc}\!\Big( \frac{1 - 2 u v}{\sqrt{v}} \Big) + \text{erfc}\!\Big( \frac{1 + 2 u v}{\sqrt{v}} \Big) \Big],
\end{equation}
where $u = (p - p_c) r$ with $p_c = 1/2$, $v = \tau / r^2$, and $\text{erfc}(x) = \frac{2}{\sqrt{\pi}} \int_x^{\infty} e^{-s^2} \, ds$ is the complementary error function. 

Fig.~\ref{fig:CDP_full}(c) shows $I(A:C|B)$ as a function of time at $p = p_c = 0.5$. For any fixed $r$, the CMI increases as $\tau$ increases. Furthermore, data for different values of $r$ collapse perfectly under the scaling with $z = 2$ and $\alpha = 0$, in agreement with the analytical result in Eq.~\eqref{Eq:supp_CMI_asymp} [see also Fig.~\ref{fig:CDP_full}(d)]. Fig.~\ref{fig:CDP_full}(e) shows $I(A:C|B)$ as a function of the error rate $p$ for different $r$, where we fix $t/r^{z} = 6$ with $z = 2$. Similar to Fig.~\ref{fig:CDP_full}(c), data for different $r$ collapse perfectly with $\nu_{\perp} = 1$ and $\alpha = 0$, again in excellent agreement with the analytical prediction of Eq.~\eqref{Eq:supp_CMI_asymp} [see Fig.~\ref{fig:CDP_full}(d)].

\subsubsection*{Strong-to-weak spontaneously symmetry breaking of the domain-wall parity}

For concreteness, we assume open boundary conditions in this section. Recall that we restrict ourselves to even times and therefore we only need to specify the spin configuration on the odd sites. Since a domain wall is defined by the condition $z_{2j-1} z_{2j+1} = -1$, the only way to change the domain-wall parity is through a \textit{nonlocal} spin-flip action such as
\begin{equation}
\label{Eq:supp_charged}
X_{L,j} \equiv \prod_{-N \le k \le j} X_{2k-1}, \qquad 
X_{R,j+1} \equiv \prod_{j \le k \le N} X_{2k+1}.
\end{equation}
We now show that the long-time state (in the same spatiotemporal limit as mentioned in the main text) $\rho(t) = \mathcal{E}^t [\rho_0]$, with $\rho_0$ defined in Eq.~\eqref{Eq:suppp_rho0_CDP}, can be understood as exhibiting strong-to-weak spontaneous symmetry breaking (SWSSB) of the domain-wall parity. As we will see, the nonlocal spin-flip operators in Eq.~\eqref{Eq:supp_charged} play a crucial role in this interpretation.

Consider time $\tau = t/2 \leq N$. Since the domain walls cannot be created out of a uniform configuration (recall $q=0$), and the initial state has a single domain wall located at the origin, the domain-wall parity cannot change. This domain-wall parity conservation can be formally expressed as 
$\mathrm{Tr}\!\left( \rho \sum_j \frac{I - Z_{2j-1} Z_{2j+1}}{2} \right) = 1 \pmod{2}$, 
with the corresponding generator
\begin{equation}
\eta 
= e^{i \pi \sum_{j=-N}^{N} (I - Z_{2j-1} Z_{2j+1})} 
= \prod_{j=-N}^{N} Z_{2j-1} Z_{2j+1} 
= Z_{-2N-1} Z_{2N+1},
\end{equation}
which is an unconventional symmetry operator in the sense that it does not act extensively on the Hilbert space. 
To see that $\eta$ acts as a strong symmetry of the mixed state, recall that a weak symmetry satisfies $U \rho U^\dagger = \rho$, whereas a strong symmetry is defined by $U \rho = e^{i\theta} \rho$. The conservation of domain-wall parity, together with the fact that the initial state contains a single domain wall, implies $
\eta \rho(t) = e^{i\pi} \rho(t) = -\rho(t)$.
Thus, $\eta$ is indeed a strong symmetry of $\rho(t)$ for $\tau = t/2 \leq N$. 

Before showing that $\lim_{t \rightarrow \infty}\lim_{N \rightarrow \infty} \rho(t = 2\tau)$ is an SWSSB state, let us first consider a concrete example by restricting to the case $p = 1$ with $\tau = N$, which provides useful intuition. In this limit, it is straightforward to see that the mixed state takes the form
\begin{equation}
\begin{aligned}
\sigma 
&= \frac{1}{2}\Big( |\uparrow_{-2N-1}\downarrow_{-2N+1}\cdots\downarrow_{2N+1}\rangle\langle\uparrow_{-2N-1}\downarrow_{-2N+1}\cdots\downarrow_{2N+1}|  \\
& \quad \quad \quad\quad\quad+ |\downarrow_{-2N-1}\cdots\downarrow_{2N-1}\uparrow_{2N+1}\rangle\langle\downarrow_{-2N-1}\cdots\downarrow_{2N-1}\uparrow_{2N+1}| \Big)\\
&\propto (I - \eta)\, \prod_{j=-N+1}^{N-1} (I - Z_{2j+1}).
\end{aligned}
\end{equation}
The fact that $\sigma$ is proportional to the projector $(I - \eta)$ already indicates that it is the fixed-point state of an SWSSB phase.
To make this precise, recall that an SWSSB state $\sigma$ is characterized by the absence of physical long-range order,
 $
\mathrm{Tr}(\sigma\, O_j O_k)_c \equiv 
\mathrm{Tr}(\sigma\, O_j O_k) - \mathrm{Tr}(\sigma\, O_j)\,\mathrm{Tr}(\sigma\, O_k)$, while possessing a non-trivial fidelity correlator 
$F(\sigma, O_j O_k\, \sigma\, O_j^\dagger O_k^\dagger)$, where $O_j$ and $O_k$ are operators charged under the symmetry \cite{lessa2024strong}. For classical mixed states $A$ and $B$, the fidelity is $F(A,B) = \sum_{\{z_j\}} \sqrt{A_{\{z_j\}} B_{\{z_j\}}}$.
In our system, the natural charged operators are 
$X_{L,-N} \equiv X_{-2N-1}$ and $X_{R,N} \equiv X_{2N+1}$, as defined in Eq.~\eqref{Eq:supp_charged}. A direct calculation shows that 
$\mathrm{Tr}(\sigma X_{L,-N}) = \mathrm{Tr}(\sigma X_{R,N}) = \mathrm{Tr}(\sigma X_{L,-N} X_{R,N}) = 0$, and thus $\sigma$ exhibits no physical long-range order.
On the other hand, one finds 
$\sigma = X_{L,-N} X_{R,N}\, \sigma\, X_{L,-N} X_{R,N}$, 
which immediately implies 
$F(\sigma, X_{L,-N} X_{R,N}\, \sigma\, X_{L,-N} X_{R,N}) = 1$. 
Therefore, $\sigma$ is indeed an SWSSB state associated with the domain-wall parity conservation.

Given the above intuition, we now show that $\lim_{t \rightarrow \infty}\lim_{N \rightarrow \infty} \rho(t = 2\tau)$ can be interpreted as an SWSSB state for any value of $p$. For the tripartition $A = [-2N-1, \cdots, -2r]$, $B = [-2r+1,\cdots,2r-1]$, and $C = [2r, \cdots, 2N+1]$, we define the following generalized fidelity correlator and demonstrate that
\begin{equation}
\label{Eq:supp_generalized_fidelity}
F(A:C|B) \equiv \sum_{j \ge r} F(\rho, X_{L,-j} X_{R,j} \rho X_{L,-j} X_{R,j}) = \frac{I(A:C|B)}{\log 2}.
\end{equation}
The operators $X_{L,-j}$ and $X_{R,j}$ defined in Eq.~\eqref{Eq:supp_charged} are charged under the domain-wall symmetry generator $\eta = Z_{-2N -1} Z_{2N +1}$. The quantity $F(A:C|B)$ therefore corresponds to summing over all independent charge-fidelity correlators supported in regions $A$ and $C$, and naturally serves as the order parameter for the SWSSB phase associated with $\eta$. Moreover, since $\lim_{r \rightarrow \infty} \lim_{\tau \rightarrow \infty} I(A:C|B) = \log 2$, it follows that $\lim_{r \rightarrow \infty} \lim_{\tau \rightarrow \infty} F(A:C|B) = 1$.

To verify Eq.~\eqref{Eq:supp_generalized_fidelity}, it suffices to show that
\begin{equation}
\label{Eq:supp_singleF}
F(\rho, X_{L,-j} X_{R,j} \rho X_{L,-j} X_{R,j}) = \Pi(\mathrm{dw} = 2j) + \Pi(\mathrm{dw} = -2j),
\end{equation}
so that $F(A:C|B) = \sum_{j \ge r} F(\rho, X_{L,-j} X_{R,j} \rho X_{L,-j} X_{R,j}) = \sum_{b \notin S_{B,\mathrm{dw}}} \Pi(B = b)$.
Eq.~\eqref{Eq:supp_singleF} follows from noting that when $\rho$ and $X_{L,-j} X_{R,j} \rho X_{L,-j} X_{R,j}$ are expanded in the domain-wall basis, they have nonzero overlap only when there is a single domain wall and is located at $\pm 2j$. Furthermore, the probability that the domain wall lies at $\pm 2j$ is the same for both $\rho$ and $X_{L,-j} X_{R,j} \rho X_{L,-j} X_{R,j}$, so that $\sqrt{\rho_{\mathrm{dw} = \pm 2j} (X_{L,-j} X_{R,j} \rho X_{L,-j} X_{R,j})_{\mathrm{dw} = \pm 2j}} = \sqrt{\rho_{\mathrm{dw} = \pm 2j}^2} = \Pi(\mathrm{dw} = \pm 2j)$. This completes the proof of Eq.~\eqref{Eq:supp_generalized_fidelity}.

\subsection{Even Number of Domain Walls}

\begin{figure*}
\centering
\includegraphics[width=\linewidth]{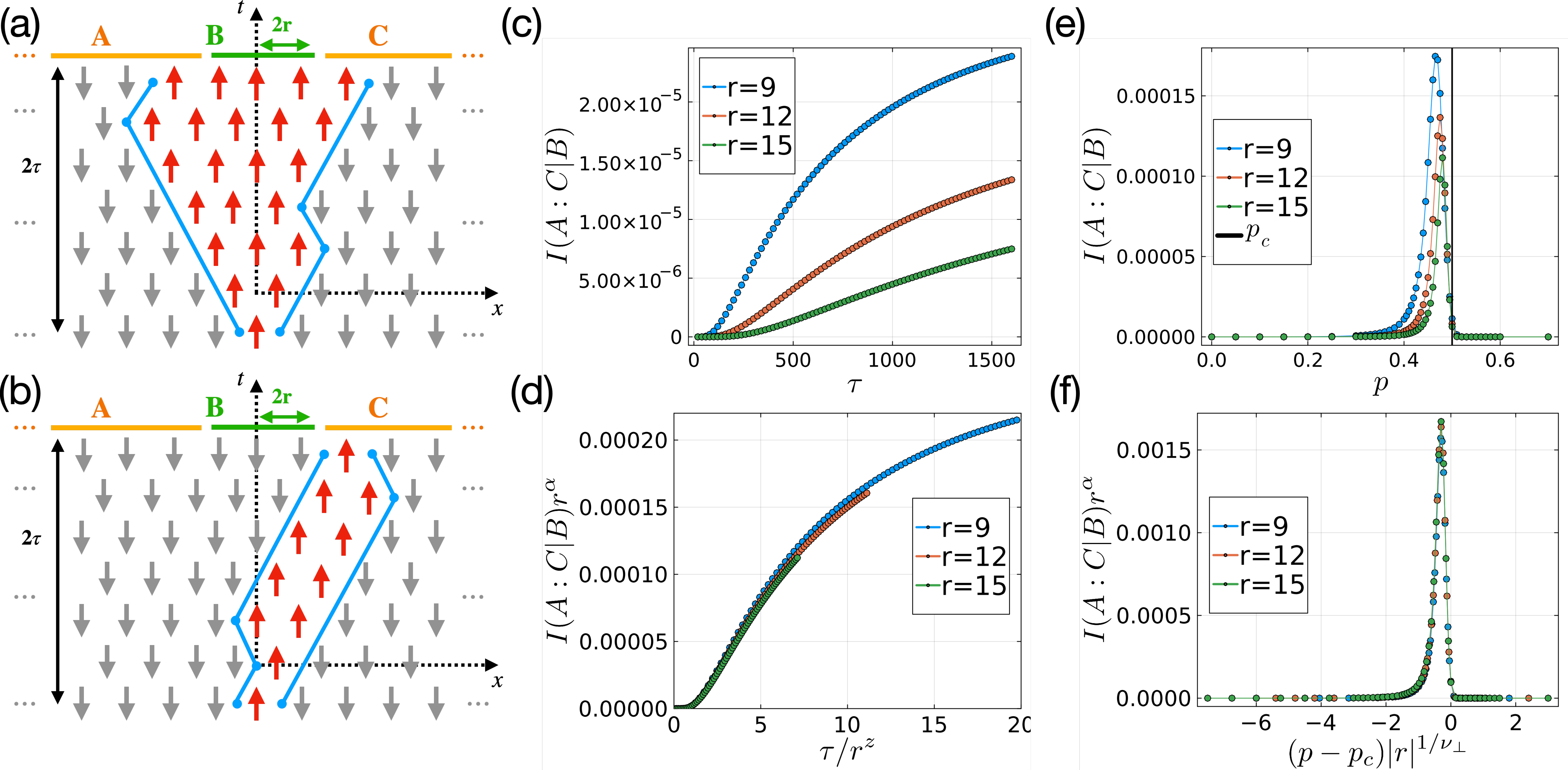}
\caption{Conditional mutual information (CMI) analysis of the compact directed percolation model with the initial state 
$|\rho_{-1}\rangle = |\cdots, \downarrow_{-4}, \downarrow_{-2}, \uparrow_{0}, \downarrow_{2}, \downarrow_{4}, \cdots\rangle$. 
One representative trajectory of $b = |\uparrow\rangle^{|B|}$ and  $b = |\downarrow\rangle^{|B|}$ is shown in (a) and (b), respectively.
(c) $I(A:C|B)$ as a function of time at $p = p_c = 0.5$. 
(d) Finite-size, finite-time scaling collapse with $z = 2$ and $\alpha = 1$. 
(e) $I(A:C|B)$ as a function of the error rate $p$ for different sizes $r$, where we fix $\tau/r^z = 6$ with $z = 2$. 
(f) Finite-size, finite-time scaling collapse with $\nu_{\perp} = 1$ and $\alpha = 1$.
}
\label{fig:CDP_full_even}
\end{figure*}

We now consider the situation where the initial state contains an even number of domain walls. For convenience, we specify the initial state at $t = -1$, which has a single active ($\uparrow$) site localized at the origin:
\begin{equation}
\label{Eq:rho_m1_even}
|\rho_{-1}\rangle = |\cdots, \downarrow_{-4}, \downarrow_{-2}, \uparrow_{0}, \downarrow_{2}, \downarrow_{4}, \cdots\rangle.
\end{equation}
We study the CMI of the state $|\rho(2\tau-1)\rangle = \mathcal{E}^{2\tau}|\rho_{-1}\rangle$ with $A = [-\infty, \cdots, -2r-1]$, $B = [-2r, \cdots, 2r]$, and $C = [2r+1, \cdots, \infty]$. See Fig.~\ref{fig:CDP_full_even}(a) for an example with $r = 2$ and $\tau = 3$.

In this case, we do not have a simple mapping in general. However, the CMI calculation can still be simplified dramatically as follows. Because domain walls can never be created and annihilate pairwise upon meeting, the full information of any trajectory is encoded entirely in the position of the left domain wall, denoted $2L-1$, and the right domain wall, denoted $2R+1$. Therefore, the mixed state is fully characterized by the absorbing probability $\rho_{\emptyset}$ and the probability distribution $\rho_{L,R}$, where $\rho_{L,R} = 0$ whenever $L > R$. Furthermore, because a domain wall cannot move faster than speed $v = 1$, at time $t = 2\tau - 1$ one must have $\rho_{L,R} = 0$ whenever $L \notin [-\tau,\tau]$ or $R \notin [-\tau,\tau]$.
In summary, at time $t = 2\tau - 1$ the mixed state $|\rho(2\tau - 1)\rangle$ can be compactly represented in terms of the absorbing probability $\rho_{\emptyset}$ and the $(2\tau + 1)\times(2\tau + 1)$ upper-triangular matrix $\rho_{L,R}$ as
\begin{equation}
\label{Eq:supp_corr}
|\rho\rangle = \rho_{\emptyset}|\emptyset\rangle 
+ \sum_{L=-\tau}^{\tau}\sum_{R=L}^{\tau} \rho_{L,R} |L,R\rangle,\qquad
\rho_{\emptyset},\ \rho_{L,R} \geq 0,\qquad 
\rho_{\emptyset} + \sum_{L=-\tau}^{\tau}\sum_{R=L}^{\tau} \rho_{L,R} = 1.
\end{equation}

Before computing the CMI, we first discuss how to obtain $\rho_{\emptyset}(t)$ and $\rho_{L,R}(t)$ as functions of time $t$. Given $\rho_{L,R}(t-1)$, which is a $(t+1)\times (t+1)$ matrix, one obtains $\rho_{L,R}(t)$, a $(t+2)\times(t+2)$ matrix, as follows. We begin by setting $\rho_{L,R}(t) = 0$ for all $L,R$ and $\rho_{\emptyset}(t) = \rho_{\emptyset}(t-1)$. 
For each $j \in [-(t+1)/2, -(t-1)/2, \cdots, (t-1)/2, (t+1)/2]$ and each $k \in [j, j+1, \cdots, (t+1)/2]$, we consider four possible updates corresponding to the shifts $(j \pm 1/2,\, k \pm 1/2)$. For the case $(j + 1/2,\, k - 1/2)$, we add $p^2 \,\rho_{L=j,\,R=k}(t-1)$ to $\rho_{L=j+1/2,\,R=k-1/2}(t)$ if $j + 1 \leq k$; otherwise we add $p^2\,\rho_{L=j,\,R=k}(t-1)$ to $\rho_{\emptyset}(t)$. 
For the remaining three update cases, $\rho_{\emptyset}(t)$ is not modified, and $\rho_{L,R}(t)$ is updated according to the corresponding probabilities. For example, for the case $(j + 1/2,\, k + 1/2)$, we add $(1-p)p\,\rho_{L=j,\,R=k}(t-1)$ to $\rho_{L=j+1/2,\, R=k+1/2}(t)$.
Combining this iteration rule with the initial condition $\rho_{\emptyset}(-1) = 0$ and $\rho_{L=0,\,R=0}(-1) = 1$ (recall that $\rho_{L,R}(t)$ is a $(t+2)\times(t+2)$ matrix, so $\rho_{L,R}(-1)$ is just a single number), one can determine $\rho_{\emptyset}(t)$ and $\rho_{L,R}(t)$ for any time $t$.

Given $\rho_{\emptyset}(t)$ and $\rho_{L,R}(t)$ for $t = 2\tau - 1$ with $\tau$ a non-negative integer, we can now compute $I(A:C|B)$ for the tripartition $A = [-\infty, \cdots, -2r-1]$, $B = [-2r, \cdots, 2r]$, and $C = [2r+1, \cdots, \infty]$. Similar to the case with an odd number of domain walls, only the configurations $b = |\uparrow\rangle^{|B|}$ and $b = |\downarrow\rangle^{|B|}$ (i.e., $b \notin S_{B,\mathrm{dw}}$) contribute to $I(A:C|B)$. Therefore,
\begin{equation}
I(A:C|B)
= \Pi(|\uparrow\rangle^{|B|})\, I(A:C \,|\, |\uparrow\rangle^{|B|})
+ \Pi(|\downarrow\rangle^{|B|})\, I(A:C \,|\, |\downarrow\rangle^{|B|}).
\end{equation}
We first consider the case $b = |\uparrow\rangle^{|B|}$. This configuration immediately implies that the left domain wall lies in region $A$ while the right domain wall lies in region $C$. Hence,
$L \in [-\tau, -r]$ and $R \in [ r, \tau]$, which leads to
\begin{equation}
\label{Eq:supp_even_up}
\begin{aligned}
\Pi(|\uparrow\rangle^{|B|})\, I(A:C \,|\, |\uparrow\rangle^{|B|})
&= \sum_{a,c} \Pi(a, |\uparrow\rangle^{|B|}, c)
    \log\Bigg(
    \frac{
        \Pi(a,|\uparrow\rangle^{|B|},c)\,
        \Pi(|\uparrow\rangle^{|B|})
    }{
        \Pi(a,|\uparrow\rangle^{|B|})\,
        \Pi(|\uparrow\rangle^{|B|},c)
    }
    \Bigg) \\
&=\sum_{L=-\tau}^{-r} \sum_{R= r}^{\tau}
    \rho_{L,R}\,
    \log\Bigg(
        \frac{
            \rho_{L,R}\;
            \big(\sum_{L''=-\tau}^{-r} \sum_{R''=r}^{\tau} \rho_{L'',R''}\big)
        }{
            \big(\sum_{L'=-\tau}^{-r} \rho_{L',R}\big)\;
            \big(\sum_{R'=r}^{\tau} \rho_{L,R'}\big)
        }
    \Bigg).
\end{aligned}
\end{equation}
We now turn to $b = |\downarrow\rangle^{|B|}$. Define $W_A$ ($W_C$) as the probability that both domain walls lie in region $A$ ($C$). By reflection symmetry $x \rightarrow -x$, one has $W_A = W_C \equiv W$, where
\begin{equation}
W = \sum_{L=-\tau}^{-r-1} \sum_{R=L}^{-r-1} \rho_{L,R}.
\end{equation}
It then follows that the marginal probability of $b = |\uparrow\rangle^{|B|}$ is
$\Pi(|\uparrow\rangle^{|B|}) = \rho_{\emptyset} + 2W$, and similarly
$\Pi(|\uparrow\rangle^{|A|},|\uparrow\rangle^{|B|})
= \Pi(|\uparrow\rangle^{|B|},|\uparrow\rangle^{|C|})
= \rho_{\emptyset} + W$.
A straightforward calculation yields
\begin{equation}
\label{Eq:supp_even_dn}
\begin{aligned}
\Pi(|\downarrow\rangle^{|B|})\, I(A:C \,|\, |\downarrow\rangle^{|B|})
&= \sum_{a,c} \Pi(a, |\downarrow\rangle^{|B|}, c)
  \log\Bigg(
    \frac{
        \Pi(a,|\downarrow\rangle^{|B|},c)\,
        \Pi(|\downarrow\rangle^{|B|})
    }{
        \Pi(a,|\downarrow\rangle^{|B|})\,
        \Pi(|\downarrow\rangle^{|B|},c)
    }
  \Bigg) \\
&= \rho_{\emptyset}
    \log\Bigg(
        \frac{
            \rho_{\emptyset}\, (\rho_{\emptyset} + 2W)
        }{
            (\rho_{\emptyset}+W)^2
        }
    \Bigg)
  + 2\sum_{L=-\tau}^{-r-1} \sum_{R=L}^{-r-1}
        \rho_{L,R}
        \log\Bigg(
            \frac{ \rho_{L,R}\, (\rho_{\emptyset}+2W) }
                 { \rho_{L,R}\, (\rho_{\emptyset}+W) }
        \Bigg) \\
&= \rho_{\emptyset}
    \log\Bigg(
        \frac{\rho_{\emptyset}(\rho_{\emptyset}+2W)}{(\rho_{\emptyset}+W)^2}
    \Bigg)
  + 2W\,
    \log\Bigg(
        \frac{\rho_{\emptyset}+2W}{\rho_{\emptyset}+W}
    \Bigg).
\end{aligned}
\end{equation}
Combining Eq.~\eqref{Eq:supp_even_up} and Eq.~\eqref{Eq:supp_even_dn} gives the total contribution to the CMI. {Our code implementation is publicly available at Ref.\footnote{See \url{https://github.com/yuhsuehchen/compact-directed-percolation-cmi} for the code implementation in Julia.}. 

Fig.~\ref{fig:CDP_full_even}(c) shows $I(A:C|B)$ as a function of time up to $t = 1600$ at $p = p_c = 0.5$. We find that data for different values of $r$ collapse well under the scaling ansatz
$I(A:C|B) = r^{-\alpha} f_1(t r^{-z})$
with $z = 2$ and $\alpha = 1$ 
[see Fig.~\ref{fig:CDP_full_even}(d)]. 
Fig.~\ref{fig:CDP_full_even}(e) shows $I(A:C|B)$ as a function of the error rate $p$ for different $r$, where we fix $t/r^{z} = 6$ with $z = 2$. Similar to Fig.~\ref{fig:CDP_full_even}(c), the data for different $r$ collapse well under the ansatz
$I(A:C|B) = r^{-\alpha} f_2\!\big((p - p_c) r^{1/\nu_{\perp}}\big)$
with $\nu_{\perp} = 1$ and $\alpha = 1$
[see Fig.~\ref{fig:CDP_full_even}(d)].

\section{Magnetization along the $p = q$ line.}

The goal of this appendix is to provide numerical evidence that along the $p =q$ line of the Domany-Kinzel model, the local magnetization converges to the steady-state value exponentially fast when $p \neq p_c \approx 0.295$.
Fig.\ref{fig:magnetization}(a) shows the local magnetization $\langle Z \rangle = \tr(\rho Z_j)$ as a function of time $t$. We consider the mixed state $\rho(t)$ to have converged to the steady state $\rho(\infty)$ when $|\langle Z(t_f)\rangle- \langle Z(t_f-1)\rangle|<10^{-10}$, at which point the time evolution is terminated. We find that when  $p$ is far away from $p_c \approx 0.295$ (i.e., $p = 0.2$ and $p = 0.4$ in Fig.\ref{fig:magnetization}(a)), $\langle Z(t) \rangle$ relaxed exponentially fast to the long-time magnetization. On the other hand, when $p$ is very close to $p_c$ (i.e., $p = 0.29,0.295,0.3$ in the Fig.\ref{fig:magnetization}(a)), $\langle Z(t) \rangle$  converges very slowly. This is manifested more transparently from when plotting $\log(\langle Z(t)\rangle - Z_\infty)$ as a function of time $t$ [see Fig.\ref{fig:magnetization}(b)], where one can see the almost perfect linear line when  $p$ is far away from $p_c \approx 0.295$. Note that we define $Z_\infty = \langle Z_{\tau}\rangle$ with $\tau = \min(400, t_f)$.

\begin{figure}
\centering
\includegraphics[width=0.6\linewidth]{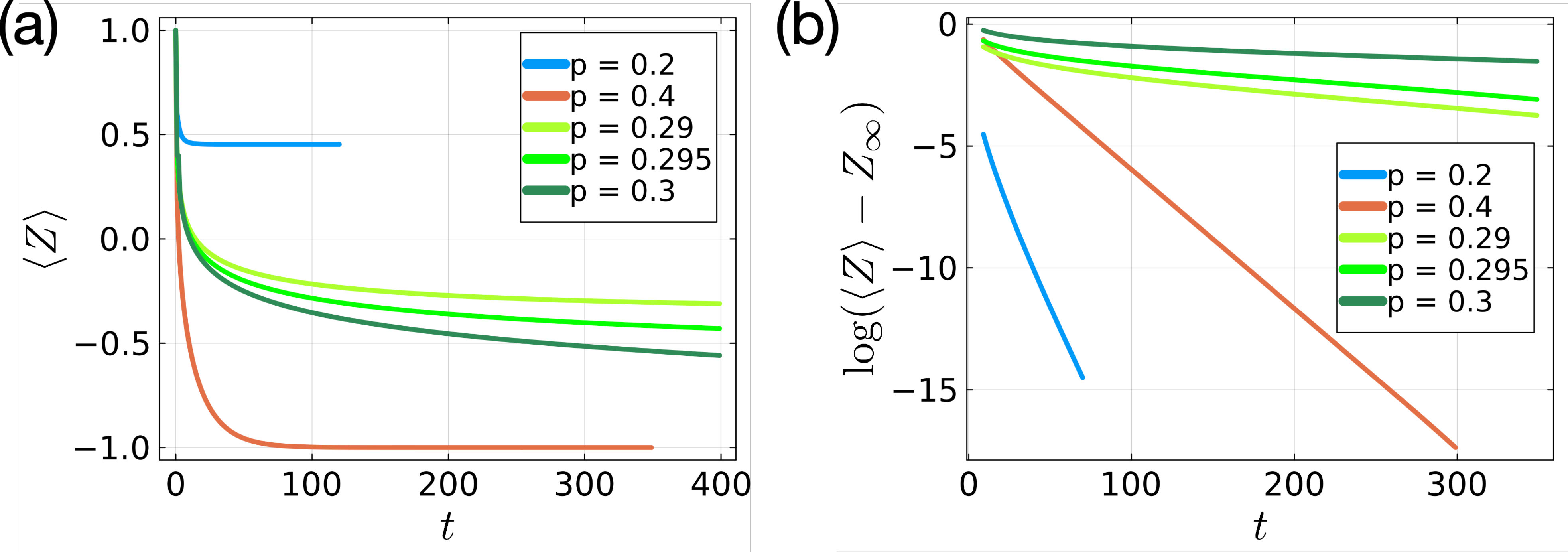}
\caption{(a) Magnetization $\langle Z\rangle $ as a function of time $t$ for different error rate $p$. Here we fix the initial state to be $\rho_0 = (|\uparrow\rangle\langle \uparrow |)^\infty$. (b) $\log(\langle Z\rangle - Z_\infty)$ as a function of time $t$. for different error rate $p$.
}
\label{fig:magnetization}
\end{figure}


\begin{thebibliography}{68}%
\makeatletter
\providecommand \@ifxundefined [1]{%
 \@ifx{#1\undefined}
}%
\providecommand \@ifnum [1]{%
 \ifnum #1\expandafter \@firstoftwo
 \else \expandafter \@secondoftwo
 \fi
}%
\providecommand \@ifx [1]{%
 \ifx #1\expandafter \@firstoftwo
 \else \expandafter \@secondoftwo
 \fi
}%
\providecommand \natexlab [1]{#1}%
\providecommand \enquote  [1]{``#1''}%
\providecommand \bibnamefont  [1]{#1}%
\providecommand \bibfnamefont [1]{#1}%
\providecommand \citenamefont [1]{#1}%
\providecommand \href@noop [0]{\@secondoftwo}%
\providecommand \href [0]{\begingroup \@sanitize@url \@href}%
\providecommand \@href[1]{\@@startlink{#1}\@@href}%
\providecommand \@@href[1]{\endgroup#1\@@endlink}%
\providecommand \@sanitize@url [0]{\catcode `\\12\catcode `\$12\catcode
  `\&12\catcode `\#12\catcode `\^12\catcode `\_12\catcode `\%12\relax}%
\providecommand \@@startlink[1]{}%
\providecommand \@@endlink[0]{}%
\providecommand \url  [0]{\begingroup\@sanitize@url \@url }%
\providecommand \@url [1]{\endgroup\@href {#1}{\urlprefix }}%
\providecommand \urlprefix  [0]{URL }%
\providecommand \Eprint [0]{\href }%
\providecommand \doibase [0]{https://doi.org/}%
\providecommand \selectlanguage [0]{\@gobble}%
\providecommand \bibinfo  [0]{\@secondoftwo}%
\providecommand \bibfield  [0]{\@secondoftwo}%
\providecommand \translation [1]{[#1]}%
\providecommand \BibitemOpen [0]{}%
\providecommand \bibitemStop [0]{}%
\providecommand \bibitemNoStop [0]{.\EOS\space}%
\providecommand \EOS [0]{\spacefactor3000\relax}%
\providecommand \BibitemShut  [1]{\csname bibitem#1\endcsname}%
\let\auto@bib@innerbib\@empty
\bibitem [{\citenamefont {Chen}\ \emph {et~al.}(2010)\citenamefont {Chen},
  \citenamefont {Gu},\ and\ \citenamefont {Wen}}]{chen2010local}%
  \BibitemOpen
  \bibfield  {author} {\bibinfo {author} {\bibfnamefont {X.}~\bibnamefont
  {Chen}}, \bibinfo {author} {\bibfnamefont {Z.-C.}\ \bibnamefont {Gu}},\ and\
  \bibinfo {author} {\bibfnamefont {X.-G.}\ \bibnamefont {Wen}},\ }\bibfield
  {title} {\bibinfo {title} {Local unitary transformation, long-range quantum
  entanglement, wave function renormalization, and topological order},\ }\href
  {https://doi.org/10.1103/PhysRevB.82.155138} {\bibfield  {journal} {\bibinfo
  {journal} {Phys. Rev. B}\ }\textbf {\bibinfo {volume} {82}},\ \bibinfo
  {pages} {155138} (\bibinfo {year} {2010})}\BibitemShut {NoStop}%
\bibitem [{\citenamefont {Hastings}\ and\ \citenamefont
  {Wen}(2005)}]{hastings2005quasiadiabatic}%
  \BibitemOpen
  \bibfield  {author} {\bibinfo {author} {\bibfnamefont {M.~B.}\ \bibnamefont
  {Hastings}}\ and\ \bibinfo {author} {\bibfnamefont {X.-G.}\ \bibnamefont
  {Wen}},\ }\bibfield  {title} {\bibinfo {title} {Quasiadiabatic continuation
  of quantum states: The stability of topological ground-state degeneracy and
  emergent gauge invariance},\ }\href@noop {} {\bibfield  {journal} {\bibinfo
  {journal} {Physical review b}\ }\textbf {\bibinfo {volume} {72}},\ \bibinfo
  {pages} {045141} (\bibinfo {year} {2005})}\BibitemShut {NoStop}%
\bibitem [{\citenamefont {Hastings}(2011)}]{hastings2011topological}%
  \BibitemOpen
  \bibfield  {author} {\bibinfo {author} {\bibfnamefont {M.~B.}\ \bibnamefont
  {Hastings}},\ }\bibfield  {title} {\bibinfo {title} {Topological order at
  nonzero temperature},\ }\href
  {https://doi.org/10.1103%2Fphysrevlett.107.210501} {\bibfield  {journal}
  {\bibinfo  {journal} {Physical review letters}\ }\textbf {\bibinfo {volume}
  {107}},\ \bibinfo {pages} {210501} (\bibinfo {year} {2011})}\BibitemShut
  {NoStop}%
\bibitem [{\citenamefont {K\"onig}\ and\ \citenamefont
  {Pastawski}(2014)}]{konig2014generating}%
  \BibitemOpen
  \bibfield  {author} {\bibinfo {author} {\bibfnamefont {R.}~\bibnamefont
  {K\"onig}}\ and\ \bibinfo {author} {\bibfnamefont {F.}~\bibnamefont
  {Pastawski}},\ }\bibfield  {title} {\bibinfo {title} {Generating topological
  order: No speedup by dissipation},\ }\href
  {https://doi.org/10.1103/PhysRevB.90.045101} {\bibfield  {journal} {\bibinfo
  {journal} {Phys. Rev. B}\ }\textbf {\bibinfo {volume} {90}},\ \bibinfo
  {pages} {045101} (\bibinfo {year} {2014})}\BibitemShut {NoStop}%
\bibitem [{\citenamefont {Coser}\ and\ \citenamefont
  {P{\'e}rez-Garc{\'\i}a}(2019)}]{coser2019classification}%
  \BibitemOpen
  \bibfield  {author} {\bibinfo {author} {\bibfnamefont {A.}~\bibnamefont
  {Coser}}\ and\ \bibinfo {author} {\bibfnamefont {D.}~\bibnamefont
  {P{\'e}rez-Garc{\'\i}a}},\ }\bibfield  {title} {\bibinfo {title}
  {Classification of phases for mixed states via fast dissipative evolution},\
  }\href {http://dx.doi.org/10.22331/q-2019-08-12-174} {\bibfield  {journal}
  {\bibinfo  {journal} {Quantum}\ }\textbf {\bibinfo {volume} {3}},\ \bibinfo
  {pages} {174} (\bibinfo {year} {2019})}\BibitemShut {NoStop}%
\bibitem [{\citenamefont {Sang}\ \emph {et~al.}(2024)\citenamefont {Sang},
  \citenamefont {Zou},\ and\ \citenamefont {Hsieh}}]{sang2024mixed}%
  \BibitemOpen
  \bibfield  {author} {\bibinfo {author} {\bibfnamefont {S.}~\bibnamefont
  {Sang}}, \bibinfo {author} {\bibfnamefont {Y.}~\bibnamefont {Zou}},\ and\
  \bibinfo {author} {\bibfnamefont {T.~H.}\ \bibnamefont {Hsieh}},\ }\bibfield
  {title} {\bibinfo {title} {Mixed-state quantum phases: Renormalization and
  quantum error correction},\ }\href
  {https://doi.org/10.1103/PhysRevX.14.031044} {\bibfield  {journal} {\bibinfo
  {journal} {Phys. Rev. X}\ }\textbf {\bibinfo {volume} {14}},\ \bibinfo
  {pages} {031044} (\bibinfo {year} {2024})}\BibitemShut {NoStop}%
\bibitem [{\citenamefont {Sang}\ and\ \citenamefont
  {Hsieh}(2025)}]{sang2025stability}%
  \BibitemOpen
  \bibfield  {author} {\bibinfo {author} {\bibfnamefont {S.}~\bibnamefont
  {Sang}}\ and\ \bibinfo {author} {\bibfnamefont {T.~H.}\ \bibnamefont
  {Hsieh}},\ }\bibfield  {title} {\bibinfo {title} {Stability of mixed-state
  quantum phases via finite markov length},\ }\href
  {https://doi.org/10.1103/PhysRevLett.134.070403} {\bibfield  {journal}
  {\bibinfo  {journal} {Phys. Rev. Lett.}\ }\textbf {\bibinfo {volume} {134}},\
  \bibinfo {pages} {070403} (\bibinfo {year} {2025})}\BibitemShut {NoStop}%
\bibitem [{\citenamefont {Sang}\ \emph {et~al.}(2025)\citenamefont {Sang},
  \citenamefont {Lessa}, \citenamefont {Mong}, \citenamefont {Grover},
  \citenamefont {Wang},\ and\ \citenamefont {Hsieh}}]{sang2025mixed}%
  \BibitemOpen
  \bibfield  {author} {\bibinfo {author} {\bibfnamefont {S.}~\bibnamefont
  {Sang}}, \bibinfo {author} {\bibfnamefont {L.~A.}\ \bibnamefont {Lessa}},
  \bibinfo {author} {\bibfnamefont {R.~S.}\ \bibnamefont {Mong}}, \bibinfo
  {author} {\bibfnamefont {T.}~\bibnamefont {Grover}}, \bibinfo {author}
  {\bibfnamefont {C.}~\bibnamefont {Wang}},\ and\ \bibinfo {author}
  {\bibfnamefont {T.~H.}\ \bibnamefont {Hsieh}},\ }\bibfield  {title} {\bibinfo
  {title} {Mixed-state phases from local reversibility},\ }\href
  {https://arxiv.org/abs/2507.02292} {\bibfield  {journal} {\bibinfo  {journal}
  {arXiv preprint arXiv:2507.02292}\ } (\bibinfo {year} {2025})}\BibitemShut
  {NoStop}%
\bibitem [{\citenamefont {Schl{\"o}gl}(1972)}]{schlogl1972chemical}%
  \BibitemOpen
  \bibfield  {author} {\bibinfo {author} {\bibfnamefont {F.}~\bibnamefont
  {Schl{\"o}gl}},\ }\bibfield  {title} {\bibinfo {title} {Chemical reaction
  models for non-equilibrium phase transitions},\ }\href@noop {} {\bibfield
  {journal} {\bibinfo  {journal} {Zeitschrift f{\"u}r physik}\ }\textbf
  {\bibinfo {volume} {253}},\ \bibinfo {pages} {147} (\bibinfo {year}
  {1972})}\BibitemShut {NoStop}%
\bibitem [{\citenamefont {Cardy}\ and\ \citenamefont
  {Sugar}(1980)}]{cardy1980directed}%
  \BibitemOpen
  \bibfield  {author} {\bibinfo {author} {\bibfnamefont {J.~L.}\ \bibnamefont
  {Cardy}}\ and\ \bibinfo {author} {\bibfnamefont {R.}~\bibnamefont {Sugar}},\
  }\bibfield  {title} {\bibinfo {title} {Directed percolation and reggeon field
  theory},\ }\href@noop {} {\bibfield  {journal} {\bibinfo  {journal} {Journal
  of Physics A: Mathematical and General}\ }\textbf {\bibinfo {volume} {13}},\
  \bibinfo {pages} {L423} (\bibinfo {year} {1980})}\BibitemShut {NoStop}%
\bibitem [{\citenamefont {Grassberger}\ and\ \citenamefont
  {Sundermeyer}(1978)}]{grassberger1978reggeon}%
  \BibitemOpen
  \bibfield  {author} {\bibinfo {author} {\bibfnamefont {P.}~\bibnamefont
  {Grassberger}}\ and\ \bibinfo {author} {\bibfnamefont {K.}~\bibnamefont
  {Sundermeyer}},\ }\bibfield  {title} {\bibinfo {title} {Reggeon field theory
  and markov processes},\ }\href@noop {} {\bibfield  {journal} {\bibinfo
  {journal} {Physics Letters B}\ }\textbf {\bibinfo {volume} {77}},\ \bibinfo
  {pages} {220} (\bibinfo {year} {1978})}\BibitemShut {NoStop}%
\bibitem [{\citenamefont {Grassberger}\ and\ \citenamefont {de~la
  Torre}(1979)}]{grassberger1979reggeon}%
  \BibitemOpen
  \bibfield  {author} {\bibinfo {author} {\bibfnamefont {P.}~\bibnamefont
  {Grassberger}}\ and\ \bibinfo {author} {\bibfnamefont {A.}~\bibnamefont
  {de~la Torre}},\ }\bibfield  {title} {\bibinfo {title} {Reggeon field theory
  (schl{\"o}gl's first model) on a lattice: Monte carlo calculations of
  critical behaviour},\ }\href@noop {} {\bibfield  {journal} {\bibinfo
  {journal} {Annals of Physics}\ }\textbf {\bibinfo {volume} {122}},\ \bibinfo
  {pages} {373} (\bibinfo {year} {1979})}\BibitemShut {NoStop}%
\bibitem [{\citenamefont {Janssen}(1981)}]{janssen1981nonequilibrium}%
  \BibitemOpen
  \bibfield  {author} {\bibinfo {author} {\bibfnamefont {H.-K.}\ \bibnamefont
  {Janssen}},\ }\bibfield  {title} {\bibinfo {title} {On the nonequilibrium
  phase transition in reaction-diffusion systems with an absorbing stationary
  state},\ }\href@noop {} {\bibfield  {journal} {\bibinfo  {journal}
  {Zeitschrift f{\"u}r Physik B Condensed Matter}\ }\textbf {\bibinfo {volume}
  {42}},\ \bibinfo {pages} {151} (\bibinfo {year} {1981})}\BibitemShut
  {NoStop}%
\bibitem [{\citenamefont {Grassberger}(1981)}]{grassberger1981phase}%
  \BibitemOpen
  \bibfield  {author} {\bibinfo {author} {\bibfnamefont {P.}~\bibnamefont
  {Grassberger}},\ }\bibfield  {title} {\bibinfo {title} {On phase transitions
  in schl{\"o}gl’s second model}\ }(\bibinfo {organization} {Springer},\
  \bibinfo {year} {1981})\ pp.\ \bibinfo {pages} {262--262}\BibitemShut
  {NoStop}%
\bibitem [{\citenamefont {Kinzel}(1985)}]{kinzel1985phase}%
  \BibitemOpen
  \bibfield  {author} {\bibinfo {author} {\bibfnamefont {W.}~\bibnamefont
  {Kinzel}},\ }\bibfield  {title} {\bibinfo {title} {Phase transitions of
  cellular automata},\ }\href@noop {} {\bibfield  {journal} {\bibinfo
  {journal} {Zeitschrift f{\"u}r Physik B Condensed Matter}\ }\textbf {\bibinfo
  {volume} {58}},\ \bibinfo {pages} {229} (\bibinfo {year} {1985})}\BibitemShut
  {NoStop}%
\bibitem [{\citenamefont {Hinrichsen}(2000)}]{hinrichsen2000non}%
  \BibitemOpen
  \bibfield  {author} {\bibinfo {author} {\bibfnamefont {H.}~\bibnamefont
  {Hinrichsen}},\ }\bibfield  {title} {\bibinfo {title} {Non-equilibrium
  critical phenomena and phase transitions into absorbing states},\ }\href@noop
  {} {\bibfield  {journal} {\bibinfo  {journal} {Advances in physics}\ }\textbf
  {\bibinfo {volume} {49}},\ \bibinfo {pages} {815} (\bibinfo {year}
  {2000})}\BibitemShut {NoStop}%
\bibitem [{\citenamefont {Goldenfeld}\ and\ \citenamefont
  {Shih}(2017)}]{goldenfeld2017turbulence}%
  \BibitemOpen
  \bibfield  {author} {\bibinfo {author} {\bibfnamefont {N.}~\bibnamefont
  {Goldenfeld}}\ and\ \bibinfo {author} {\bibfnamefont {H.-Y.}\ \bibnamefont
  {Shih}},\ }\bibfield  {title} {\bibinfo {title} {Turbulence as a problem in
  non-equilibrium statistical mechanics},\ }\href@noop {} {\bibfield  {journal}
  {\bibinfo  {journal} {Journal of Statistical Physics}\ }\textbf {\bibinfo
  {volume} {167}},\ \bibinfo {pages} {575} (\bibinfo {year}
  {2017})}\BibitemShut {NoStop}%
\bibitem [{\citenamefont {Täuber}(2014)}]{tauber2014critical}%
  \BibitemOpen
  \bibfield  {author} {\bibinfo {author} {\bibfnamefont {U.~C.}\ \bibnamefont
  {Täuber}},\ }\bibfield  {title} {\bibinfo {title} {Critical dynamics: A
  field theory approach to equilibrium and non-equilibrium scaling behavior}\
  }(\bibinfo  {publisher} {Cambridge University Press},\ \bibinfo {year}
  {2014})\BibitemShut {NoStop}%
\bibitem [{\citenamefont {Cardy}(1996)}]{cardy1996scaling}%
  \BibitemOpen
  \bibfield  {author} {\bibinfo {author} {\bibfnamefont {J.}~\bibnamefont
  {Cardy}},\ }\href@noop {} {\emph {\bibinfo {title} {Scaling and
  renormalization in statistical physics}}},\ Vol.~\bibinfo {volume} {5}\
  (\bibinfo  {publisher} {Cambridge university press},\ \bibinfo {year}
  {1996})\BibitemShut {NoStop}%
\bibitem [{\citenamefont {Fan}\ \emph {et~al.}(2024)\citenamefont {Fan},
  \citenamefont {Bao}, \citenamefont {Altman},\ and\ \citenamefont
  {Vishwanath}}]{fan2023diagnostics}%
  \BibitemOpen
  \bibfield  {author} {\bibinfo {author} {\bibfnamefont {R.}~\bibnamefont
  {Fan}}, \bibinfo {author} {\bibfnamefont {Y.}~\bibnamefont {Bao}}, \bibinfo
  {author} {\bibfnamefont {E.}~\bibnamefont {Altman}},\ and\ \bibinfo {author}
  {\bibfnamefont {A.}~\bibnamefont {Vishwanath}},\ }\bibfield  {title}
  {\bibinfo {title} {Diagnostics of mixed-state topological order and breakdown
  of quantum memory},\ }\href {https://doi.org/10.1103/PRXQuantum.5.020343}
  {\bibfield  {journal} {\bibinfo  {journal} {PRX Quantum}\ }\textbf {\bibinfo
  {volume} {5}},\ \bibinfo {pages} {020343} (\bibinfo {year}
  {2024})}\BibitemShut {NoStop}%
\bibitem [{\citenamefont {Bao}\ \emph {et~al.}(2023)\citenamefont {Bao},
  \citenamefont {Fan}, \citenamefont {Vishwanath},\ and\ \citenamefont
  {Altman}}]{bao2023mixed}%
  \BibitemOpen
  \bibfield  {author} {\bibinfo {author} {\bibfnamefont {Y.}~\bibnamefont
  {Bao}}, \bibinfo {author} {\bibfnamefont {R.}~\bibnamefont {Fan}}, \bibinfo
  {author} {\bibfnamefont {A.}~\bibnamefont {Vishwanath}},\ and\ \bibinfo
  {author} {\bibfnamefont {E.}~\bibnamefont {Altman}},\ }\bibfield  {title}
  {\bibinfo {title} {Mixed-state topological order and the errorfield double
  formulation of decoherence-induced transitions},\ }\href
  {https://arxiv.org/abs/2301.05687} {\bibfield  {journal} {\bibinfo  {journal}
  {arXiv preprint arXiv:2301.05687}\ } (\bibinfo {year} {2023})}\BibitemShut
  {NoStop}%
\bibitem [{\citenamefont {Lee}\ \emph {et~al.}(2023)\citenamefont {Lee},
  \citenamefont {Jian},\ and\ \citenamefont {Xu}}]{lee2023quantum}%
  \BibitemOpen
  \bibfield  {author} {\bibinfo {author} {\bibfnamefont {J.~Y.}\ \bibnamefont
  {Lee}}, \bibinfo {author} {\bibfnamefont {C.-M.}\ \bibnamefont {Jian}},\ and\
  \bibinfo {author} {\bibfnamefont {C.}~\bibnamefont {Xu}},\ }\bibfield
  {title} {\bibinfo {title} {Quantum criticality under decoherence or weak
  measurement},\ }\href {https://doi.org/10.1103/PRXQuantum.4.030317}
  {\bibfield  {journal} {\bibinfo  {journal} {PRX Quantum}\ }\textbf {\bibinfo
  {volume} {4}},\ \bibinfo {pages} {030317} (\bibinfo {year}
  {2023})}\BibitemShut {NoStop}%
\bibitem [{\citenamefont {Wang}\ \emph
  {et~al.}(2025{\natexlab{a}})\citenamefont {Wang}, \citenamefont {Wu},\ and\
  \citenamefont {Wang}}]{wang2025intrinsic}%
  \BibitemOpen
  \bibfield  {author} {\bibinfo {author} {\bibfnamefont {Z.}~\bibnamefont
  {Wang}}, \bibinfo {author} {\bibfnamefont {Z.}~\bibnamefont {Wu}},\ and\
  \bibinfo {author} {\bibfnamefont {Z.}~\bibnamefont {Wang}},\ }\bibfield
  {title} {\bibinfo {title} {Intrinsic mixed-state topological order},\ }\href
  {https://doi.org/10.1103/PRXQuantum.6.010314} {\bibfield  {journal} {\bibinfo
   {journal} {PRX Quantum}\ }\textbf {\bibinfo {volume} {6}},\ \bibinfo {pages}
  {010314} (\bibinfo {year} {2025}{\natexlab{a}})}\BibitemShut {NoStop}%
\bibitem [{\citenamefont {Chen}\ and\ \citenamefont
  {Grover}(2024)}]{chen2023separability}%
  \BibitemOpen
  \bibfield  {author} {\bibinfo {author} {\bibfnamefont {Y.-H.}\ \bibnamefont
  {Chen}}\ and\ \bibinfo {author} {\bibfnamefont {T.}~\bibnamefont {Grover}},\
  }\bibfield  {title} {\bibinfo {title} {Separability transitions in
  topological states induced by local decoherence},\ }\href
  {https://doi.org/10.1103/PhysRevLett.132.170602} {\bibfield  {journal}
  {\bibinfo  {journal} {Phys. Rev. Lett.}\ }\textbf {\bibinfo {volume} {132}},\
  \bibinfo {pages} {170602} (\bibinfo {year} {2024})}\BibitemShut {NoStop}%
\bibitem [{\citenamefont {Li}\ and\ \citenamefont
  {Mong}(2024)}]{li2024replica}%
  \BibitemOpen
  \bibfield  {author} {\bibinfo {author} {\bibfnamefont {Z.}~\bibnamefont
  {Li}}\ and\ \bibinfo {author} {\bibfnamefont {R.~S.}\ \bibnamefont {Mong}},\
  }\bibfield  {title} {\bibinfo {title} {Replica topological order in quantum
  mixed states and quantum error correction},\ }\href@noop {} {\bibfield
  {journal} {\bibinfo  {journal} {arXiv preprint arXiv:2402.09516}\ } (\bibinfo
  {year} {2024})}\BibitemShut {NoStop}%
\bibitem [{\citenamefont {Wang}\ \emph
  {et~al.}(2025{\natexlab{b}})\citenamefont {Wang}, \citenamefont {Song},
  \citenamefont {Meng},\ and\ \citenamefont {Grover}}]{wang2025analog}%
  \BibitemOpen
  \bibfield  {author} {\bibinfo {author} {\bibfnamefont {T.-T.}\ \bibnamefont
  {Wang}}, \bibinfo {author} {\bibfnamefont {M.}~\bibnamefont {Song}}, \bibinfo
  {author} {\bibfnamefont {Z.~Y.}\ \bibnamefont {Meng}},\ and\ \bibinfo
  {author} {\bibfnamefont {T.}~\bibnamefont {Grover}},\ }\bibfield  {title}
  {\bibinfo {title} {Analog of topological entanglement entropy for mixed
  states},\ }\href {https://doi.org/10.1103/PRXQuantum.6.010358} {\bibfield
  {journal} {\bibinfo  {journal} {PRX Quantum}\ }\textbf {\bibinfo {volume}
  {6}},\ \bibinfo {pages} {010358} (\bibinfo {year}
  {2025}{\natexlab{b}})}\BibitemShut {NoStop}%
\bibitem [{\citenamefont {Lessa}\ \emph {et~al.}(2025)\citenamefont {Lessa},
  \citenamefont {Sang}, \citenamefont {Lu}, \citenamefont {Hsieh},\ and\
  \citenamefont {Wang}}]{lessa2025higher}%
  \BibitemOpen
  \bibfield  {author} {\bibinfo {author} {\bibfnamefont {L.~A.}\ \bibnamefont
  {Lessa}}, \bibinfo {author} {\bibfnamefont {S.}~\bibnamefont {Sang}},
  \bibinfo {author} {\bibfnamefont {T.-C.}\ \bibnamefont {Lu}}, \bibinfo
  {author} {\bibfnamefont {T.~H.}\ \bibnamefont {Hsieh}},\ and\ \bibinfo
  {author} {\bibfnamefont {C.}~\bibnamefont {Wang}},\ }\bibfield  {title}
  {\bibinfo {title} {Higher-form anomaly and long-range entanglement of mixed
  states},\ }\href@noop {} {\bibfield  {journal} {\bibinfo  {journal} {arXiv
  preprint arXiv:2503.12792}\ } (\bibinfo {year} {2025})}\BibitemShut {NoStop}%
\bibitem [{\citenamefont {Ellison}\ and\ \citenamefont
  {Cheng}(2025)}]{ellison2025toward}%
  \BibitemOpen
  \bibfield  {author} {\bibinfo {author} {\bibfnamefont {T.~D.}\ \bibnamefont
  {Ellison}}\ and\ \bibinfo {author} {\bibfnamefont {M.}~\bibnamefont
  {Cheng}},\ }\bibfield  {title} {\bibinfo {title} {Toward a classification of
  mixed-state topological orders in two dimensions},\ }\href
  {https://doi.org/10.1103/PRXQuantum.6.010315} {\bibfield  {journal} {\bibinfo
   {journal} {PRX Quantum}\ }\textbf {\bibinfo {volume} {6}},\ \bibinfo {pages}
  {010315} (\bibinfo {year} {2025})}\BibitemShut {NoStop}%
\bibitem [{\citenamefont {Sohal}\ and\ \citenamefont
  {Prem}(2025)}]{sohal2025noisy}%
  \BibitemOpen
  \bibfield  {author} {\bibinfo {author} {\bibfnamefont {R.}~\bibnamefont
  {Sohal}}\ and\ \bibinfo {author} {\bibfnamefont {A.}~\bibnamefont {Prem}},\
  }\bibfield  {title} {\bibinfo {title} {Noisy approach to intrinsically
  mixed-state topological order},\ }\href
  {https://doi.org/10.1103/PRXQuantum.6.010313} {\bibfield  {journal} {\bibinfo
   {journal} {PRX Quantum}\ }\textbf {\bibinfo {volume} {6}},\ \bibinfo {pages}
  {010313} (\bibinfo {year} {2025})}\BibitemShut {NoStop}%
\bibitem [{\citenamefont {Balasubramanian}\ \emph {et~al.}(2024)\citenamefont
  {Balasubramanian}, \citenamefont {Davydova},\ and\ \citenamefont
  {Lake}}]{balasubramanian2024local}%
  \BibitemOpen
  \bibfield  {author} {\bibinfo {author} {\bibfnamefont {S.}~\bibnamefont
  {Balasubramanian}}, \bibinfo {author} {\bibfnamefont {M.}~\bibnamefont
  {Davydova}},\ and\ \bibinfo {author} {\bibfnamefont {E.}~\bibnamefont
  {Lake}},\ }\bibfield  {title} {\bibinfo {title} {A local automaton for the 2d
  toric code},\ }\href@noop {} {\bibfield  {journal} {\bibinfo  {journal}
  {arXiv preprint arXiv:2412.19803}\ } (\bibinfo {year} {2024})}\BibitemShut
  {NoStop}%
\bibitem [{\citenamefont {Ohya}\ and\ \citenamefont
  {Petz}(2004)}]{ohya2004quantum}%
  \BibitemOpen
  \bibfield  {author} {\bibinfo {author} {\bibfnamefont {M.}~\bibnamefont
  {Ohya}}\ and\ \bibinfo {author} {\bibfnamefont {D.}~\bibnamefont {Petz}},\
  }\href@noop {} {\emph {\bibinfo {title} {Quantum entropy and its use}}}\
  (\bibinfo  {publisher} {Springer Science \& Business Media},\ \bibinfo {year}
  {2004})\BibitemShut {NoStop}%
\bibitem [{\citenamefont {Petz}(1986)}]{petz1986sufficient}%
  \BibitemOpen
  \bibfield  {author} {\bibinfo {author} {\bibfnamefont {D.}~\bibnamefont
  {Petz}},\ }\bibfield  {title} {\bibinfo {title} {Sufficient subalgebras and
  the relative entropy of states of a von neumann algebra},\ }\href@noop {}
  {\bibfield  {journal} {\bibinfo  {journal} {Communications in mathematical
  physics}\ }\textbf {\bibinfo {volume} {105}},\ \bibinfo {pages} {123}
  (\bibinfo {year} {1986})}\BibitemShut {NoStop}%
\bibitem [{\citenamefont {Junge}\ \emph {et~al.}(2018)\citenamefont {Junge},
  \citenamefont {Renner}, \citenamefont {Sutter}, \citenamefont {Wilde},\ and\
  \citenamefont {Winter}}]{junge2018universal}%
  \BibitemOpen
  \bibfield  {author} {\bibinfo {author} {\bibfnamefont {M.}~\bibnamefont
  {Junge}}, \bibinfo {author} {\bibfnamefont {R.}~\bibnamefont {Renner}},
  \bibinfo {author} {\bibfnamefont {D.}~\bibnamefont {Sutter}}, \bibinfo
  {author} {\bibfnamefont {M.~M.}\ \bibnamefont {Wilde}},\ and\ \bibinfo
  {author} {\bibfnamefont {A.}~\bibnamefont {Winter}},\ }\bibfield  {title}
  {\bibinfo {title} {Universal recovery maps and approximate sufficiency of
  quantum relative entropy},\ }in\ \href@noop {} {\emph {\bibinfo {booktitle}
  {Annales Henri Poincar{\'e}}}},\ Vol.~\bibinfo {volume} {19}\ (\bibinfo
  {organization} {Springer},\ \bibinfo {year} {2018})\ pp.\ \bibinfo {pages}
  {2955--2978}\BibitemShut {NoStop}%
\bibitem [{\citenamefont {Rakovszky}\ \emph {et~al.}(2024)\citenamefont
  {Rakovszky}, \citenamefont {Gopalakrishnan},\ and\ \citenamefont {von
  Keyserlingk}}]{rakovszky2024defining}%
  \BibitemOpen
  \bibfield  {author} {\bibinfo {author} {\bibfnamefont {T.}~\bibnamefont
  {Rakovszky}}, \bibinfo {author} {\bibfnamefont {S.}~\bibnamefont
  {Gopalakrishnan}},\ and\ \bibinfo {author} {\bibfnamefont {C.}~\bibnamefont
  {von Keyserlingk}},\ }\bibfield  {title} {\bibinfo {title} {Defining stable
  phases of open quantum systems},\ }\href
  {https://doi.org/10.1103/PhysRevX.14.041031} {\bibfield  {journal} {\bibinfo
  {journal} {Phys. Rev. X}\ }\textbf {\bibinfo {volume} {14}},\ \bibinfo
  {pages} {041031} (\bibinfo {year} {2024})}\BibitemShut {NoStop}%
\bibitem [{\citenamefont {Cubitt}\ \emph {et~al.}(2015)\citenamefont {Cubitt},
  \citenamefont {Lucia}, \citenamefont {Michalakis},\ and\ \citenamefont
  {Perez-Garcia}}]{cubitt2015stability}%
  \BibitemOpen
  \bibfield  {author} {\bibinfo {author} {\bibfnamefont {T.~S.}\ \bibnamefont
  {Cubitt}}, \bibinfo {author} {\bibfnamefont {A.}~\bibnamefont {Lucia}},
  \bibinfo {author} {\bibfnamefont {S.}~\bibnamefont {Michalakis}},\ and\
  \bibinfo {author} {\bibfnamefont {D.}~\bibnamefont {Perez-Garcia}},\
  }\bibfield  {title} {\bibinfo {title} {Stability of local quantum dissipative
  systems},\ }\href@noop {} {\bibfield  {journal} {\bibinfo  {journal}
  {Communications in Mathematical Physics}\ }\textbf {\bibinfo {volume}
  {337}},\ \bibinfo {pages} {1275} (\bibinfo {year} {2015})}\BibitemShut
  {NoStop}%
\bibitem [{\citenamefont {Liu}\ and\ \citenamefont
  {Lieu}(2024)}]{liu2024dissipative}%
  \BibitemOpen
  \bibfield  {author} {\bibinfo {author} {\bibfnamefont {Y.-J.}\ \bibnamefont
  {Liu}}\ and\ \bibinfo {author} {\bibfnamefont {S.}~\bibnamefont {Lieu}},\
  }\bibfield  {title} {\bibinfo {title} {Dissipative phase transitions and
  passive error correction},\ }\href@noop {} {\bibfield  {journal} {\bibinfo
  {journal} {Physical Review A}\ }\textbf {\bibinfo {volume} {109}},\ \bibinfo
  {pages} {022422} (\bibinfo {year} {2024})}\BibitemShut {NoStop}%
\bibitem [{\citenamefont {Clifford}\ and\ \citenamefont
  {Hammersley}(1971)}]{clifford1971markov}%
  \BibitemOpen
  \bibfield  {author} {\bibinfo {author} {\bibfnamefont {P.}~\bibnamefont
  {Clifford}}\ and\ \bibinfo {author} {\bibfnamefont {J.~M.}\ \bibnamefont
  {Hammersley}},\ }\bibfield  {title} {\bibinfo {title} {Markov fields on
  finite graphs and lattices}} (\bibinfo {year} {1971}),\ \bibinfo {note}
  {manuscript, University of Oxford}\BibitemShut {NoStop}%
\bibitem [{\citenamefont {Leifer}\ and\ \citenamefont
  {Poulin}(2008)}]{leifer2008quantum}%
  \BibitemOpen
  \bibfield  {author} {\bibinfo {author} {\bibfnamefont {M.~S.}\ \bibnamefont
  {Leifer}}\ and\ \bibinfo {author} {\bibfnamefont {D.}~\bibnamefont
  {Poulin}},\ }\bibfield  {title} {\bibinfo {title} {Quantum graphical models
  and belief propagation},\ }\href@noop {} {\bibfield  {journal} {\bibinfo
  {journal} {Annals of Physics}\ }\textbf {\bibinfo {volume} {323}},\ \bibinfo
  {pages} {1899} (\bibinfo {year} {2008})}\BibitemShut {NoStop}%
\bibitem [{\citenamefont {Brown}\ and\ \citenamefont
  {Poulin}(2012)}]{brown2012quantum}%
  \BibitemOpen
  \bibfield  {author} {\bibinfo {author} {\bibfnamefont {W.}~\bibnamefont
  {Brown}}\ and\ \bibinfo {author} {\bibfnamefont {D.}~\bibnamefont {Poulin}},\
  }\bibfield  {title} {\bibinfo {title} {Quantum markov networks and commuting
  hamiltonians},\ }\href@noop {} {\bibfield  {journal} {\bibinfo  {journal}
  {arXiv preprint arXiv:1206.0755}\ } (\bibinfo {year} {2012})}\BibitemShut
  {NoStop}%
\bibitem [{\citenamefont {Chen}\ and\ \citenamefont
  {Rouz{\'e}}(2025)}]{chen2025quantum}%
  \BibitemOpen
  \bibfield  {author} {\bibinfo {author} {\bibfnamefont {C.-F.}\ \bibnamefont
  {Chen}}\ and\ \bibinfo {author} {\bibfnamefont {C.}~\bibnamefont
  {Rouz{\'e}}},\ }\bibfield  {title} {\bibinfo {title} {Quantum gibbs states
  are locally markovian},\ }\href@noop {} {\bibfield  {journal} {\bibinfo
  {journal} {arXiv preprint arXiv:2504.02208}\ } (\bibinfo {year}
  {2025})}\BibitemShut {NoStop}%
\bibitem [{\citenamefont {Lebowitz}\ and\ \citenamefont
  {Schonmann}(1988)}]{lebowitz1988pseudo}%
  \BibitemOpen
  \bibfield  {author} {\bibinfo {author} {\bibfnamefont {J.}~\bibnamefont
  {Lebowitz}}\ and\ \bibinfo {author} {\bibfnamefont {R.~H.}\ \bibnamefont
  {Schonmann}},\ }\bibfield  {title} {\bibinfo {title} {Pseudo-free energies
  and large deviations for non gibbsian fkg measures},\ }\href@noop {}
  {\bibfield  {journal} {\bibinfo  {journal} {Probability theory and related
  fields}\ }\textbf {\bibinfo {volume} {77}},\ \bibinfo {pages} {49} (\bibinfo
  {year} {1988})}\BibitemShut {NoStop}%
\bibitem [{\citenamefont {Lebowitz}\ \emph {et~al.}(1990)\citenamefont
  {Lebowitz}, \citenamefont {Maes},\ and\ \citenamefont
  {Speer}}]{lebowitz1990statistical}%
  \BibitemOpen
  \bibfield  {author} {\bibinfo {author} {\bibfnamefont {J.~L.}\ \bibnamefont
  {Lebowitz}}, \bibinfo {author} {\bibfnamefont {C.}~\bibnamefont {Maes}},\
  and\ \bibinfo {author} {\bibfnamefont {E.~R.}\ \bibnamefont {Speer}},\
  }\bibfield  {title} {\bibinfo {title} {Statistical mechanics of probabilistic
  cellular automata},\ }\href@noop {} {\bibfield  {journal} {\bibinfo
  {journal} {Journal of statistical physics}\ }\textbf {\bibinfo {volume}
  {59}},\ \bibinfo {pages} {117} (\bibinfo {year} {1990})}\BibitemShut
  {NoStop}%
\bibitem [{\citenamefont {Van~Enter}\ \emph {et~al.}(1993)\citenamefont
  {Van~Enter}, \citenamefont {Fern{\'a}ndez},\ and\ \citenamefont
  {Sokal}}]{van1993regularity}%
  \BibitemOpen
  \bibfield  {author} {\bibinfo {author} {\bibfnamefont {A.~C.}\ \bibnamefont
  {Van~Enter}}, \bibinfo {author} {\bibfnamefont {R.}~\bibnamefont
  {Fern{\'a}ndez}},\ and\ \bibinfo {author} {\bibfnamefont {A.~D.}\
  \bibnamefont {Sokal}},\ }\bibfield  {title} {\bibinfo {title} {Regularity
  properties and pathologies of position-space renormalization-group
  transformations: Scope and limitations of gibbsian theory},\ }\href@noop {}
  {\bibfield  {journal} {\bibinfo  {journal} {Journal of Statistical Physics}\
  }\textbf {\bibinfo {volume} {72}},\ \bibinfo {pages} {879} (\bibinfo {year}
  {1993})}\BibitemShut {NoStop}%
\bibitem [{\citenamefont {Fern{\'a}ndez}\ and\ \citenamefont
  {Toom}(2001)}]{fernandez2001non}%
  \BibitemOpen
  \bibfield  {author} {\bibinfo {author} {\bibfnamefont {R.}~\bibnamefont
  {Fern{\'a}ndez}}\ and\ \bibinfo {author} {\bibfnamefont {A.}~\bibnamefont
  {Toom}},\ }\bibfield  {title} {\bibinfo {title} {Non-gibbsianness of the
  invariant measures of non-reversible cellular automata with totally
  asymmetric noise},\ }\href@noop {} {\bibfield  {journal} {\bibinfo  {journal}
  {arXiv preprint math-ph/0101014}\ } (\bibinfo {year} {2001})}\BibitemShut
  {NoStop}%
\bibitem [{\citenamefont {Domany}\ and\ \citenamefont
  {Kinzel}(1984)}]{domany1984equivalence}%
  \BibitemOpen
  \bibfield  {author} {\bibinfo {author} {\bibfnamefont {E.}~\bibnamefont
  {Domany}}\ and\ \bibinfo {author} {\bibfnamefont {W.}~\bibnamefont
  {Kinzel}},\ }\bibfield  {title} {\bibinfo {title} {Equivalence of cellular
  automata to ising models and directed percolation},\ }\href@noop {}
  {\bibfield  {journal} {\bibinfo  {journal} {Physical review letters}\
  }\textbf {\bibinfo {volume} {53}},\ \bibinfo {pages} {311} (\bibinfo {year}
  {1984})}\BibitemShut {NoStop}%
\bibitem [{\citenamefont {Essam}(1989)}]{essam1989directed}%
  \BibitemOpen
  \bibfield  {author} {\bibinfo {author} {\bibfnamefont {J.}~\bibnamefont
  {Essam}},\ }\bibfield  {title} {\bibinfo {title} {Directed compact
  percolation: cluster size and hyperscaling},\ }\href@noop {} {\bibfield
  {journal} {\bibinfo  {journal} {Journal of Physics A: Mathematical and
  General}\ }\textbf {\bibinfo {volume} {22}},\ \bibinfo {pages} {4927}
  (\bibinfo {year} {1989})}\BibitemShut {NoStop}%
\bibitem [{\citenamefont {Janssen}(2005)}]{janssen2005survival}%
  \BibitemOpen
  \bibfield  {author} {\bibinfo {author} {\bibfnamefont {H.-K.}\ \bibnamefont
  {Janssen}},\ }\bibfield  {title} {\bibinfo {title} {Survival and percolation
  probabilities in the field theory of growth models},\ }\href@noop {}
  {\bibfield  {journal} {\bibinfo  {journal} {Journal of Physics: Condensed
  Matter}\ }\textbf {\bibinfo {volume} {17}},\ \bibinfo {pages} {S1973}
  (\bibinfo {year} {2005})}\BibitemShut {NoStop}%
\bibitem [{\citenamefont {Stavskaja}\ and\ \citenamefont
  {Pjatetskii-Shapiro}(1968)}]{stavskaja1968homogeneous}%
  \BibitemOpen
  \bibfield  {author} {\bibinfo {author} {\bibfnamefont {O.}~\bibnamefont
  {Stavskaja}}\ and\ \bibinfo {author} {\bibfnamefont {I.}~\bibnamefont
  {Pjatetskii-Shapiro}},\ }\bibfield  {title} {\bibinfo {title} {Homogeneous
  networks of spontaneously active elements},\ }\href@noop {} {\bibfield
  {journal} {\bibinfo  {journal} {Problemy Kibernet}\ }\textbf {\bibinfo
  {volume} {20}},\ \bibinfo {pages} {91} (\bibinfo {year} {1968})}\BibitemShut
  {NoStop}%
\bibitem [{\citenamefont {Janssen}\ \emph {et~al.}(1988)\citenamefont
  {Janssen}, \citenamefont {Schaub},\ and\ \citenamefont
  {Schmittmann}}]{janssen1988finite}%
  \BibitemOpen
  \bibfield  {author} {\bibinfo {author} {\bibfnamefont {H.}~\bibnamefont
  {Janssen}}, \bibinfo {author} {\bibfnamefont {B.}~\bibnamefont {Schaub}},\
  and\ \bibinfo {author} {\bibfnamefont {B.}~\bibnamefont {Schmittmann}},\
  }\bibfield  {title} {\bibinfo {title} {Finite size scaling for directed
  percolation and related stochastic evolution processes},\ }\href@noop {}
  {\bibfield  {journal} {\bibinfo  {journal} {Zeitschrift f{\"u}r Physik B
  Condensed Matter}\ }\textbf {\bibinfo {volume} {71}},\ \bibinfo {pages} {377}
  (\bibinfo {year} {1988})}\BibitemShut {NoStop}%
\bibitem [{\citenamefont {L{\"u}beck}\ and\ \citenamefont
  {Janssen}(2005)}]{lubeck2005finite}%
  \BibitemOpen
  \bibfield  {author} {\bibinfo {author} {\bibfnamefont {S.}~\bibnamefont
  {L{\"u}beck}}\ and\ \bibinfo {author} {\bibfnamefont {H.-K.}\ \bibnamefont
  {Janssen}},\ }\bibfield  {title} {\bibinfo {title} {Finite-size scaling of
  directed percolation above the upper critical dimension},\ }\href@noop {}
  {\bibfield  {journal} {\bibinfo  {journal} {Physical Review E—Statistical,
  Nonlinear, and Soft Matter Physics}\ }\textbf {\bibinfo {volume} {72}},\
  \bibinfo {pages} {016119} (\bibinfo {year} {2005})}\BibitemShut {NoStop}%
\bibitem [{Note1()}]{Note1}%
  \BibitemOpen
  \bibinfo {note} {Note that this probability distribution is not normalizable
  --- one can introduce an infinitesimal magnetic field to make it
  normalizable, see Ref.~\cite {lubeck2005finite}}\BibitemShut {NoStop}%
\bibitem [{sup()}]{supplement}%
  \BibitemOpen
  \href@noop {} {\bibinfo  {journal} {See Supplemental Material for details.}\
  }\BibitemShut {NoStop}%
\bibitem [{\citenamefont {Blythe}\ and\ \citenamefont
  {Evans}(2007)}]{blythe2007nonequilibrium}%
  \BibitemOpen
\bibfield  {journal} {  }\bibfield  {author} {\bibinfo {author} {\bibfnamefont
  {R.~A.}\ \bibnamefont {Blythe}}\ and\ \bibinfo {author} {\bibfnamefont
  {M.~R.}\ \bibnamefont {Evans}},\ }\bibfield  {title} {\bibinfo {title}
  {Nonequilibrium steady states of matrix-product form: a solver's guide},\
  }\href@noop {} {\bibfield  {journal} {\bibinfo  {journal} {Journal of Physics
  A: Mathematical and Theoretical}\ }\textbf {\bibinfo {volume} {40}},\
  \bibinfo {pages} {R333} (\bibinfo {year} {2007})}\BibitemShut {NoStop}%
\bibitem [{\citenamefont {Temme}\ and\ \citenamefont
  {Verstraete}(2010)}]{temme2010stochastic}%
  \BibitemOpen
  \bibfield  {author} {\bibinfo {author} {\bibfnamefont {K.}~\bibnamefont
  {Temme}}\ and\ \bibinfo {author} {\bibfnamefont {F.}~\bibnamefont
  {Verstraete}},\ }\bibfield  {title} {\bibinfo {title} {Stochastic matrix
  product states},\ }\href@noop {} {\bibfield  {journal} {\bibinfo  {journal}
  {Physical review letters}\ }\textbf {\bibinfo {volume} {104}},\ \bibinfo
  {pages} {210502} (\bibinfo {year} {2010})}\BibitemShut {NoStop}%
\bibitem [{\citenamefont {Vidal}(2007)}]{vidal2007classical}%
  \BibitemOpen
  \bibfield  {author} {\bibinfo {author} {\bibfnamefont {G.}~\bibnamefont
  {Vidal}},\ }\bibfield  {title} {\bibinfo {title} {Classical simulation of
  infinite-size quantum lattice systems in one spatial dimension},\ }\href@noop
  {} {\bibfield  {journal} {\bibinfo  {journal} {Physical review letters}\
  }\textbf {\bibinfo {volume} {98}},\ \bibinfo {pages} {070201} (\bibinfo
  {year} {2007})}\BibitemShut {NoStop}%
\bibitem [{Note2()}]{Note2}%
  \BibitemOpen
  \bibinfo {note} {See \protect \url
  {https://github.com/yuhsuehchen/directed-percolation-cmi} for the code
  implementation in Julia.}\BibitemShut {Stop}%
\bibitem [{\citenamefont {Aldous}(2006)}]{aldous2006random}%
  \BibitemOpen
  \bibfield  {author} {\bibinfo {author} {\bibfnamefont {D.}~\bibnamefont
  {Aldous}},\ }\bibfield  {title} {\bibinfo {title} {Random walks on finite
  groups and rapidly mixing markov chains},\ }in\ \href@noop {} {\emph
  {\bibinfo {booktitle} {S{\'e}minaire de Probabilit{\'e}s XVII 1981/82:
  Proceedings}}}\ (\bibinfo  {publisher} {Springer},\ \bibinfo {year} {2006})\
  pp.\ \bibinfo {pages} {243--297}\BibitemShut {NoStop}%
\bibitem [{\citenamefont {de~Maere}\ and\ \citenamefont
  {Ponselet}(2012)}]{de2012exponential}%
  \BibitemOpen
  \bibfield  {author} {\bibinfo {author} {\bibfnamefont {A.}~\bibnamefont
  {de~Maere}}\ and\ \bibinfo {author} {\bibfnamefont {L.}~\bibnamefont
  {Ponselet}},\ }\bibfield  {title} {\bibinfo {title} {Exponential decay of
  correlations for strongly coupled toom probabilistic cellular automata},\
  }\href@noop {} {\bibfield  {journal} {\bibinfo  {journal} {Journal of
  Statistical Physics}\ }\textbf {\bibinfo {volume} {147}},\ \bibinfo {pages}
  {634} (\bibinfo {year} {2012})}\BibitemShut {NoStop}%
\bibitem [{\citenamefont {Ferris}\ and\ \citenamefont
  {Vidal}(2012)}]{ferris2012perfect}%
  \BibitemOpen
  \bibfield  {author} {\bibinfo {author} {\bibfnamefont {A.~J.}\ \bibnamefont
  {Ferris}}\ and\ \bibinfo {author} {\bibfnamefont {G.}~\bibnamefont {Vidal}},\
  }\bibfield  {title} {\bibinfo {title} {Perfect sampling with unitary tensor
  networks},\ }\href@noop {} {\bibfield  {journal} {\bibinfo  {journal}
  {Physical Review B—Condensed Matter and Materials Physics}\ }\textbf
  {\bibinfo {volume} {85}},\ \bibinfo {pages} {165146} (\bibinfo {year}
  {2012})}\BibitemShut {NoStop}%
\bibitem [{\citenamefont {Lloyd}\ \emph {et~al.}(2025)\citenamefont {Lloyd},
  \citenamefont {Abanin},\ and\ \citenamefont
  {Gopalakrishnan}}]{lloyd2025diverging}%
  \BibitemOpen
  \bibfield  {author} {\bibinfo {author} {\bibfnamefont {J.}~\bibnamefont
  {Lloyd}}, \bibinfo {author} {\bibfnamefont {D.~A.}\ \bibnamefont {Abanin}},\
  and\ \bibinfo {author} {\bibfnamefont {S.}~\bibnamefont {Gopalakrishnan}},\
  }\bibfield  {title} {\bibinfo {title} {Diverging conditional correlation
  lengths in the approach to high temperature},\ }\href@noop {} {\bibfield
  {journal} {\bibinfo  {journal} {arXiv preprint arXiv:2508.02567}\ } (\bibinfo
  {year} {2025})}\BibitemShut {NoStop}%
\bibitem [{\citenamefont {Mendon{\c{c}}a}(2011)}]{mendoncca2011monte}%
  \BibitemOpen
  \bibfield  {author} {\bibinfo {author} {\bibfnamefont {J.~R.~G.}\
  \bibnamefont {Mendon{\c{c}}a}},\ }\bibfield  {title} {\bibinfo {title} {Monte
  carlo investigation of the critical behavior of stavskaya’s probabilistic
  cellular automaton},\ }\href@noop {} {\bibfield  {journal} {\bibinfo
  {journal} {Physical Review E—Statistical, Nonlinear, and Soft Matter
  Physics}\ }\textbf {\bibinfo {volume} {83}},\ \bibinfo {pages} {012102}
  (\bibinfo {year} {2011})}\BibitemShut {NoStop}%
\bibitem [{\citenamefont {Lessa}\ \emph {et~al.}(2024)\citenamefont {Lessa},
  \citenamefont {Ma}, \citenamefont {Zhang}, \citenamefont {Bi}, \citenamefont
  {Cheng},\ and\ \citenamefont {Wang}}]{lessa2024strong}%
  \BibitemOpen
  \bibfield  {author} {\bibinfo {author} {\bibfnamefont {L.~A.}\ \bibnamefont
  {Lessa}}, \bibinfo {author} {\bibfnamefont {R.}~\bibnamefont {Ma}}, \bibinfo
  {author} {\bibfnamefont {J.-H.}\ \bibnamefont {Zhang}}, \bibinfo {author}
  {\bibfnamefont {Z.}~\bibnamefont {Bi}}, \bibinfo {author} {\bibfnamefont
  {M.}~\bibnamefont {Cheng}},\ and\ \bibinfo {author} {\bibfnamefont
  {C.}~\bibnamefont {Wang}},\ }\bibfield  {title} {\bibinfo {title}
  {Strong-to-weak spontaneous symmetry breaking in mixed quantum states},\
  }\href {https://arxiv.org/abs/2405.03639} {\bibfield  {journal} {\bibinfo
  {journal} {arXiv preprint arXiv:2405.03639}\ } (\bibinfo {year}
  {2024})}\BibitemShut {NoStop}%
\bibitem [{\citenamefont {Kardar}\ \emph {et~al.}(1986)\citenamefont {Kardar},
  \citenamefont {Parisi},\ and\ \citenamefont {Zhang}}]{kardar1986dynamic}%
  \BibitemOpen
  \bibfield  {author} {\bibinfo {author} {\bibfnamefont {M.}~\bibnamefont
  {Kardar}}, \bibinfo {author} {\bibfnamefont {G.}~\bibnamefont {Parisi}},\
  and\ \bibinfo {author} {\bibfnamefont {Y.-C.}\ \bibnamefont {Zhang}},\
  }\bibfield  {title} {\bibinfo {title} {Dynamic scaling of growing
  interfaces},\ }\href {https://doi.org/10.1103/PhysRevLett.56.889} {\bibfield
  {journal} {\bibinfo  {journal} {Physical Review Letters}\ }\textbf {\bibinfo
  {volume} {56}},\ \bibinfo {pages} {889} (\bibinfo {year} {1986})}\BibitemShut
  {NoStop}%
\bibitem [{\citenamefont {Toner}\ and\ \citenamefont
  {Tu}(1995)}]{toner1995long}%
  \BibitemOpen
  \bibfield  {author} {\bibinfo {author} {\bibfnamefont {J.}~\bibnamefont
  {Toner}}\ and\ \bibinfo {author} {\bibfnamefont {Y.}~\bibnamefont {Tu}},\
  }\bibfield  {title} {\bibinfo {title} {Long-range order in a two-dimensional
  dynamical {XY} model: How birds fly together},\ }\href
  {https://doi.org/10.1103/PhysRevLett.75.4326} {\bibfield  {journal} {\bibinfo
   {journal} {Physical Review Letters}\ }\textbf {\bibinfo {volume} {75}},\
  \bibinfo {pages} {4326} (\bibinfo {year} {1995})}\BibitemShut {NoStop}%
\bibitem [{\citenamefont {Toom}(1974)}]{toom1974nonergodic}%
  \BibitemOpen
  \bibfield  {author} {\bibinfo {author} {\bibfnamefont {A.~L.}\ \bibnamefont
  {Toom}},\ }\bibfield  {title} {\bibinfo {title} {Nonergodic multidimensional
  system of automata},\ }\href@noop {} {\bibfield  {journal} {\bibinfo
  {journal} {Problemy Peredachi Informatsii}\ }\textbf {\bibinfo {volume}
  {10}},\ \bibinfo {pages} {70} (\bibinfo {year} {1974})}\BibitemShut {NoStop}%
\bibitem [{\citenamefont {Toom}(1980)}]{toom1980stable}%
  \BibitemOpen
  \bibfield  {author} {\bibinfo {author} {\bibfnamefont {A.~L.}\ \bibnamefont
  {Toom}},\ }\bibfield  {title} {\bibinfo {title} {Stable and attractive
  trajectories in multicomponent systems},\ }in\ \href@noop {} {\emph {\bibinfo
  {booktitle} {Multicomponent Random Systems}}},\ \bibinfo {series} {Advances
  in Probability and Related Topics}, Vol.~\bibinfo {volume} {6},\ \bibinfo
  {editor} {edited by\ \bibinfo {editor} {\bibfnamefont {R.~L.}\ \bibnamefont
  {Dobrushin}}\ and\ \bibinfo {editor} {\bibfnamefont {Y.~G.}\ \bibnamefont
  {Sinai}}}\ (\bibinfo  {publisher} {Marcel Dekker},\ \bibinfo {address} {New
  York},\ \bibinfo {year} {1980})\ pp.\ \bibinfo {pages} {549--575}\BibitemShut
  {NoStop}%
\bibitem [{\citenamefont {He}\ \emph {et~al.}(1990)\citenamefont {He},
  \citenamefont {Jayaprakash},\ and\ \citenamefont
  {Grinstein}}]{he1990generic}%
  \BibitemOpen
  \bibfield  {author} {\bibinfo {author} {\bibfnamefont {Y.}~\bibnamefont
  {He}}, \bibinfo {author} {\bibfnamefont {C.}~\bibnamefont {Jayaprakash}},\
  and\ \bibinfo {author} {\bibfnamefont {G.}~\bibnamefont {Grinstein}},\
  }\bibfield  {title} {\bibinfo {title} {Generic nonergodic behavior in locally
  interacting continuous systems},\ }\href@noop {} {\bibfield  {journal}
  {\bibinfo  {journal} {Physical Review A}\ }\textbf {\bibinfo {volume} {42}},\
  \bibinfo {pages} {3348} (\bibinfo {year} {1990})}\BibitemShut {NoStop}%
\bibitem [{Note3()}]{Note3}%
  \BibitemOpen
  \bibinfo {note} {See \protect \url
  {https://github.com/yuhsuehchen/compact-directed-percolation-cmi} for the
  code implementation in Julia.}\BibitemShut {Stop}%
\end{thebibliography}
\end{document}